\newcommand{\publisheddoi}{10.1093/jcde/qwac076}

\documentclass[11pt,a4paper]{article}

\usepackage[utf8]{inputenc}
\usepackage[T1]{fontenc}
\usepackage{microtype}
\usepackage{graphicx}
\usepackage{epstopdf}
\usepackage{siunitx}
\usepackage{amssymb}
\usepackage{amsmath}
\usepackage{subcaption}
\usepackage{caption}
\usepackage{array}
\usepackage{float}
\usepackage{url}
\usepackage{xcolor}
\usepackage[american]{babel}
\usepackage{csquotes}
\usepackage[
  style=apa,
  sortcites=true,
  sorting=nyt,
  backend=biber,
  hyperref=true,
]{biblatex}
\DeclareLanguageMapping{american}{american-apa}
\usepackage[colorlinks,linkcolor=blue,citecolor=blue,urlcolor=blue]{hyperref}
\usepackage{authblk}

\usepackage{geometry}
\usepackage{soul}
\sethlcolor{white}
\makeatletter
\AtBeginDocument{\let\hl\@firstofone}
\makeatother
\DeclareRobustCommand{\hlcyan}[1]{#1}

\makeatletter
\renewcommand{\part}[1]{\section*{#1}}
\makeatother

\title{A geometric modelling framework to support the\\
design of heterogeneous lattice structures with\\
non-linearly varying geometry}

\author[1]{Nikita Letov}
\author[1]{Yaoyao Fiona Zhao\thanks{Corresponding author: \href{mailto:yaoyao.zhao@mcgill.ca}{yaoyao.zhao@mcgill.ca}}}
\affil[1]{Department of Mechanical Engineering, Faculty of Engineering,\\
McGill University, Montr\'eal, Qu\'ebec H3A 0G4, Canada}

\date{}

\begin{document}
\maketitle

\begin{abstract}
\noindent\textit{Published version:} \emph{Journal of Computational Design and Engineering}.\\
DOI: \url{https://doi.org/\publisheddoi}

\medskip
Geometric modelling has been a crucial component of
the design process ever since the introduction of the
first Computer-Aided Design (CAD) systems.
Additive Manufacturing (AM) pushes design freedom to
previously unachievable limits.
AM allows the manufacturing of lattice structures
which are otherwise close to impossible to be manufactured
conventionally.
Yet, the geometric modelling of heterogeneous lattice structures
is still greatly limited.
Thus, the AM industry is now in a situation where
the manufacturing capabilities exceed the geometric
modelling capabilities.
While there have been advancements in the modelling of heterogeneous
lattice structures, the review of relevant literature
revealed critical limitations of the existing approaches.
These limitations include their inability to model
non-linear variation
of geometric parameters, as well as the limited amount of controllable
geometric parameters.
This work presents a novel
geometric modelling methodology based
on function representation as an attempt to bridge this gap.
The proposed approach avoids the manual definition of geometric parameters
and provides a method to control them with mathematical functions instead.
A software prototype implementing the proposed approach is
presented, and several use-cases are analysed.
\end{abstract}


\paragraph{Key points.}
\begin{itemize}
\item Heterogeneous lattice structures can be easily 3D printed,
yet their modelling is still challenging.
\item A novel functional approach is proposed to simplify the process
of the modelling of heterogeneous lattice structures.
\item Non-linear variation of geometric parameters such as thickness and
cross-section shape is supported.
\item Beam-based and surface-based lattice structure are supported by the proposed
approach.
\item The approach is implemented in a software prototype and tested by
modelling lattice structures with functionally defined heterogeneity.
\end{itemize}

\newcommand{\subfigw}{0.43\textwidth}

\setcounter{section}{0}
\section{Introduction}
\label{sec:introduction}

Additive manufacturing (AM) has been pushing the limits of the design freedom
since its introduction~\parencite{Bikas2016, Yang2015, Yang2015b}.
AM allows a much higher geometric complexity compared to the more conventional
means of manufacturing~\parencite{Jared2017}.
The increased supported complexity of AM is mainly associated with the
subtractive nature of conventional
manufacturing contrary to AM~\parencite{ZHANG20211}.

\begin{figure*}
        \centering
        \begin{subfigure}[b]{0.3\textwidth}
                \centering
                \includegraphics[height=5cm]{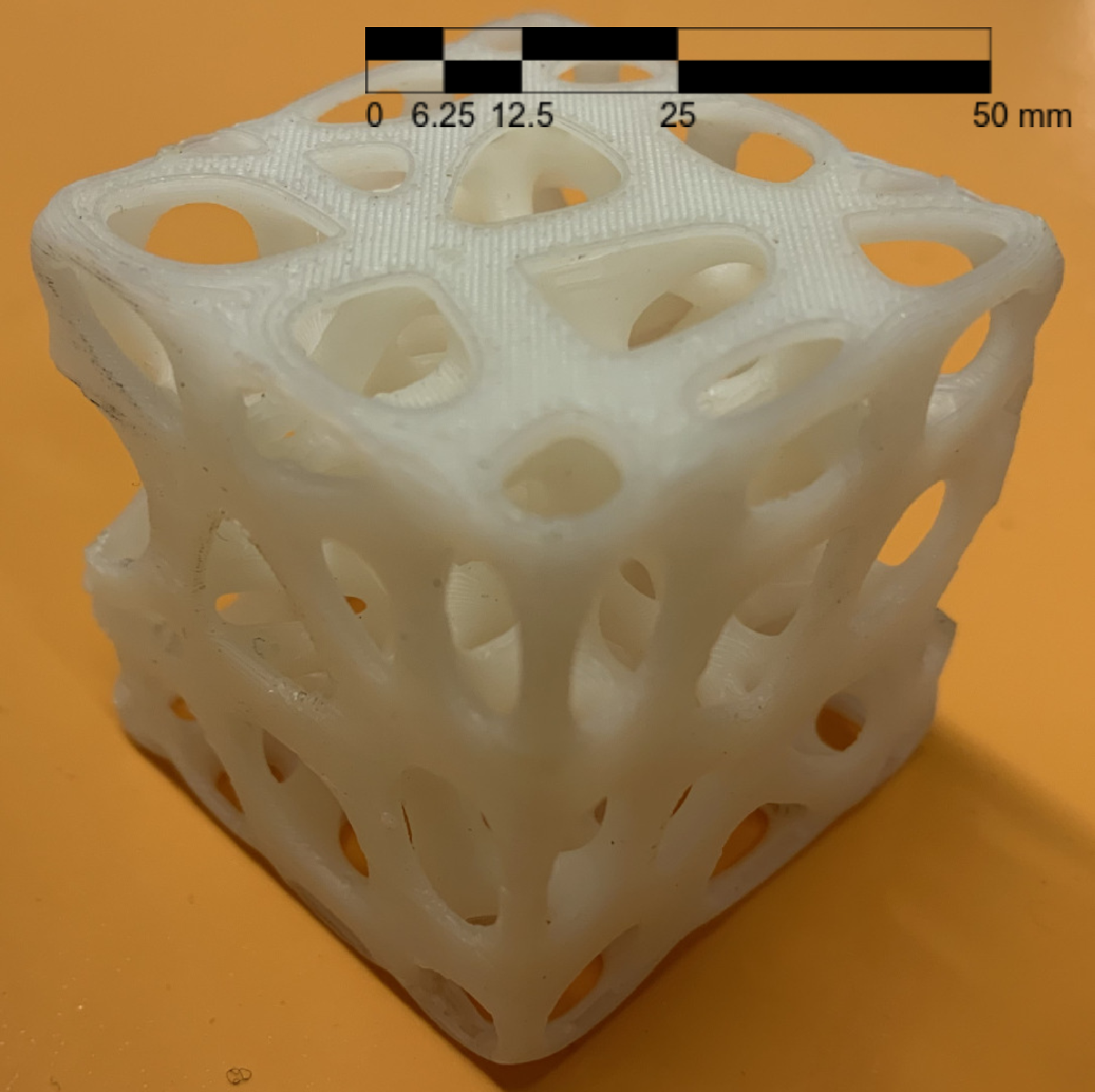}
                \caption{Voronoi topology}
                \label{fig:voronoi}
        \end{subfigure}
        \hfill
        \begin{subfigure}[b]{0.3\textwidth}
                \centering
                \includegraphics[height=5cm]{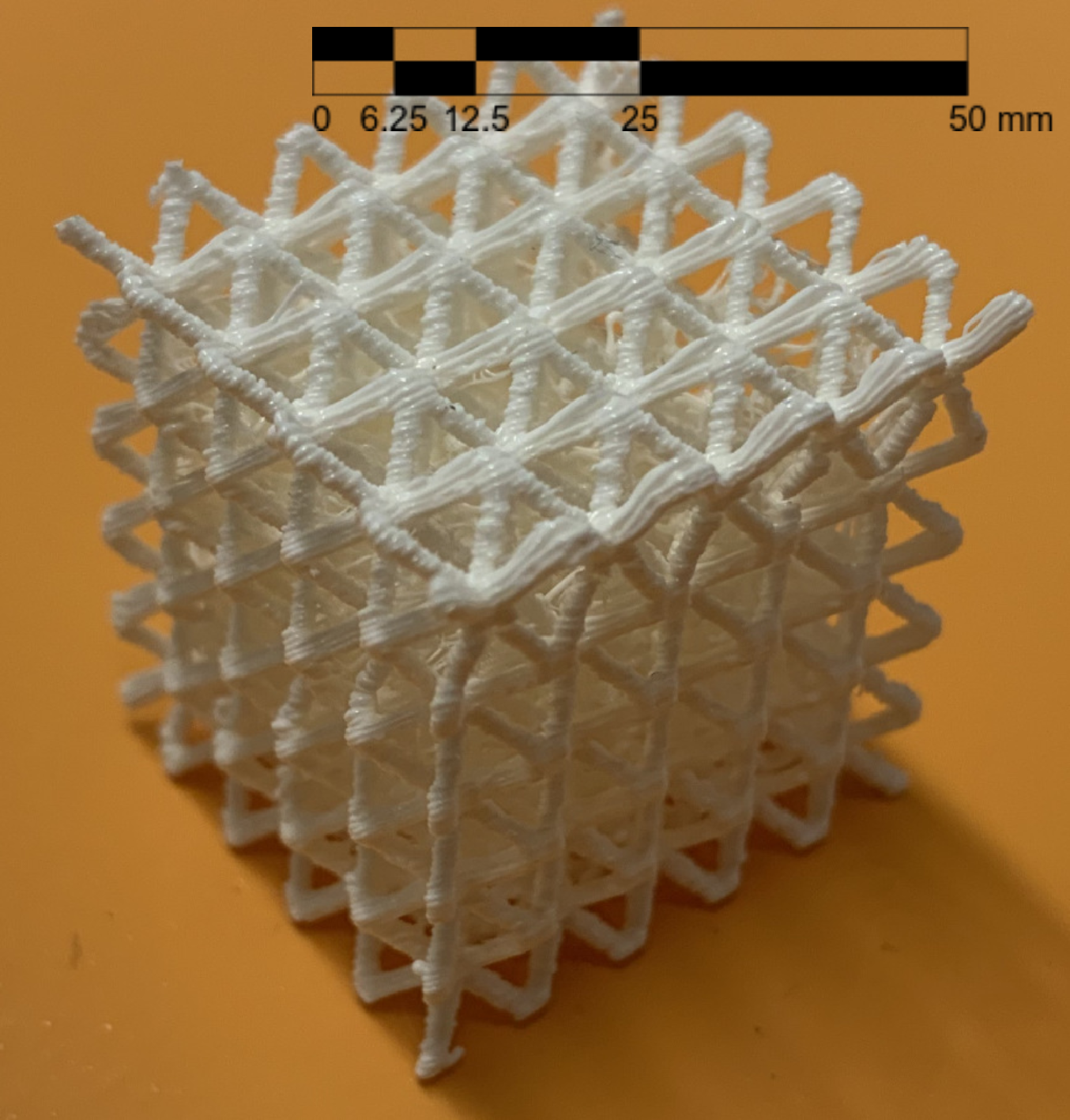}
                \caption{Body-centred cubic (BCC) topology}
                \label{fig:realbcc}
        \end{subfigure}
        \hfill
        \begin{subfigure}[b]{0.3\textwidth}
                \centering
                \includegraphics[height=5cm]{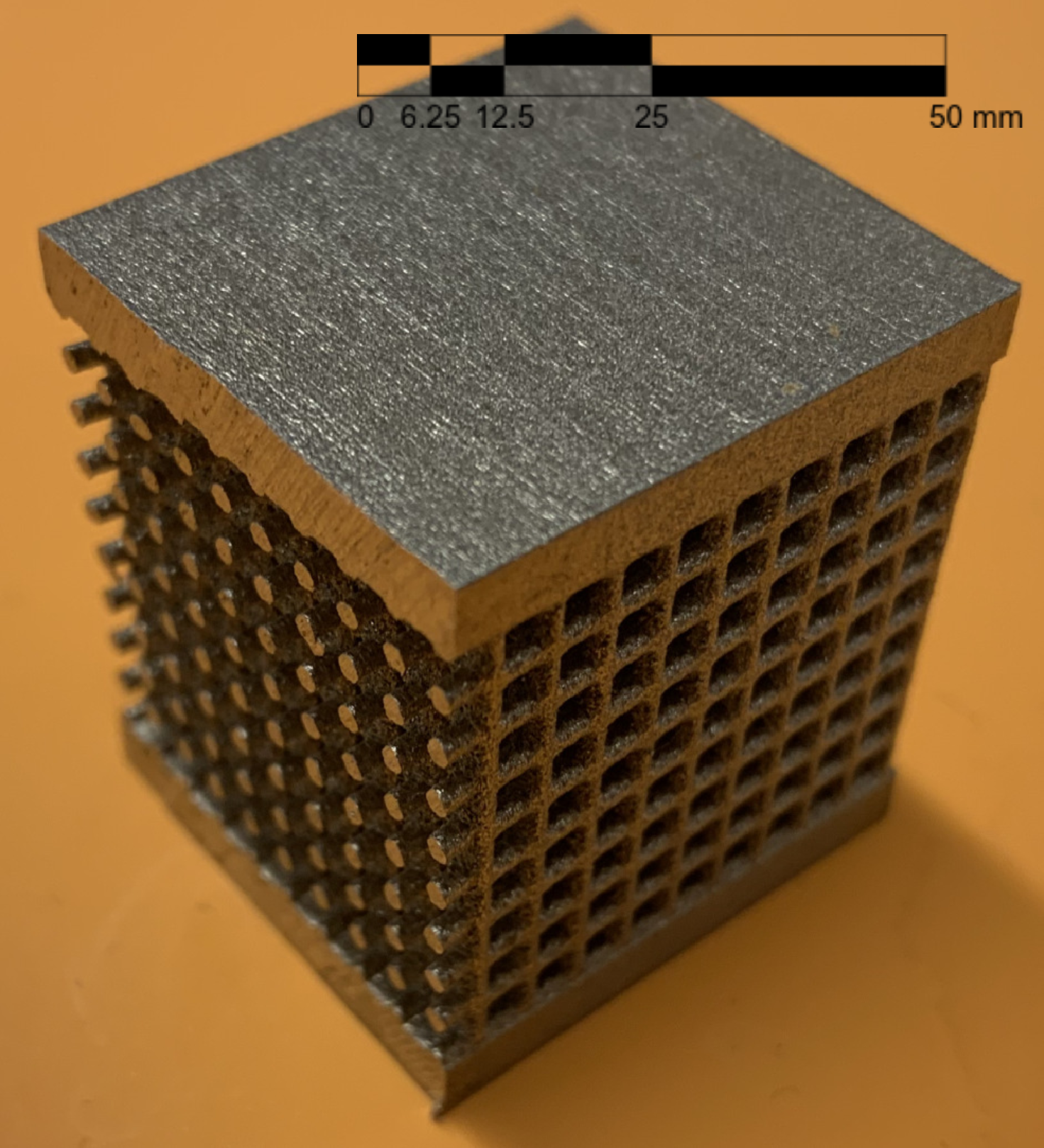}
                \caption{Simple cubic topology, metal printed}
                \label{fig:simple-cubic}
        \end{subfigure}

        \caption{Examples of lattice structures that can be
        printed with additive manufacturing}
        \label{fig:lattices}
\end{figure*}

Lattice structures provide an example of a complex shape
that can be easily manufactured by AM means while being
nearly impossible to be manufactured by conventional subtractive
manufacturing processes.
Lattice structures have found their industrial applications in
areas such as the aerospace industry, medicine, construction,
etc.~\parencite{dong2017survey, balzannikov2016usage, azarov2019composite, wang2022computational}.
They can be designed to possess unique properties which include
a reduced mass
and a negative Poisson's
ratio~\parencite{savio2017optimization, mohammadi2020hybrid}.
Fig.~\ref{fig:lattices} illustrates a few examples of lattice structures
produced by AM.

\begin{figure}
    \centering
    \includegraphics[width=0.45\textwidth]{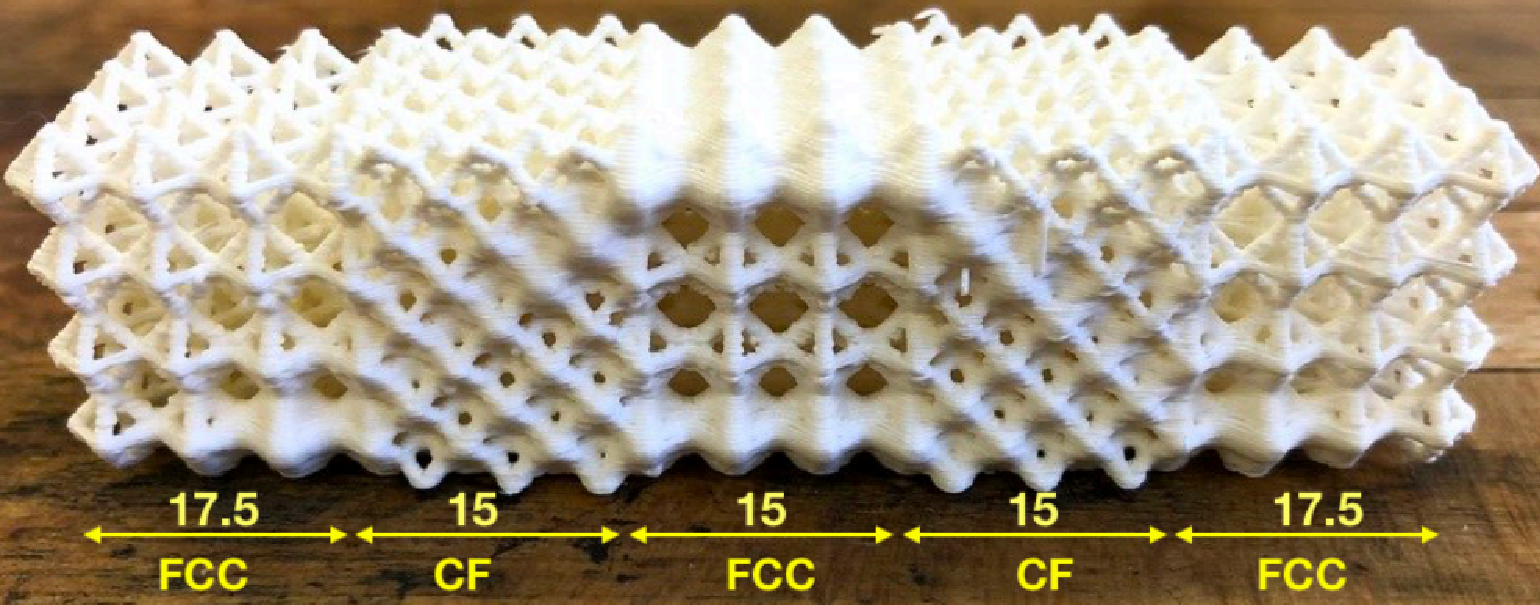}
    \caption{A printed example of a heterogeneous lattice structure
    with multiple topologies and a varying beam
    thickness~\parencite{leonardi2019additive}}
    \label{fig:hetero-example}
\end{figure}

A unique niche in AM of lattice structures is devoted to heterogeneous
lattice structures.
Heterogeneous lattice structures allow the combination of
multiple sets of geometric properties in different regions of
a single structure.
Different geometric properties enable different mechanical
properties like stress and strain, and can enhance these properties
in specific directions~\parencite{zhang2021novel}.
They are often used, for example, in
the case when a lattice structure
is subject to a varying load throughout its different
regions~\parencite{leonardi2019additive}.
Heterogeneous lattice structures have found their application in
biomedicine~\parencite{yang2021combinational},
dentistry~\parencite{javaid2019current}, vibration
management~\parencite{matlack2016composite},
bridge construction~\parencite{koltunov2021monitoring},
heat exchange~\parencite{kim20203d},
and more.
An example of a heterogeneous lattice structure is shown in
Fig.~\ref{fig:hetero-example},
which shows a lattice structure that consists of multiple topologies.

Manufacturing of heterogeneous lattice structures is nowadays
feasible with the majority of modern commercially available
3D printers that provide sufficient manufacturing
freedom~\parencite{leonardi2019additive}.
However, manufacturing freedom is not limited solely by
manufacturing capabilities and human imagination.
The question of how to design a part as a digital twin
of a conceptual design is still open~\parencite{Roam2009, Tao2019}.
In an attempt to digitise and automate the process of
creation of engineering documentation, computer-aided design (CAD)
software packages have been used extensively in the design
and manufacturing
industry~\parencite{Bezier1989, Boyer2009, tornincasa2010future}.
CAD approaches provide tools to transform the conceptual design,
which could have originated
even on the back of a napkin, into a three-dimensional (3D)
solid model~\parencite{Shah2000, Roam2009}.
These tools have been developed primarily to support conventional
manufacturing.
Existing CAD software shows significant limitations when dealing with
geometrically complex shapes such as heterogeneous lattices and
bio-inspired structures~\parencite{Liu2021}.

For example, in any feature-based CAD tool, a homogeneous lattice
structure
can be represented as a linear pattern of unit cells with cylindrical
beams positioned as edges of a cube.
However, feature-based CAD tools typically do not allow changing unit cell
parameters, such as the beam diameter, in patterns.
This significant limitation
restricts feature-based CAD to model homogeneous lattice structures
only and
makes it unfeasible to design
heterogeneous lattice structures.

\begin{figure}
    \centering
    \includegraphics[width=0.45\textwidth]{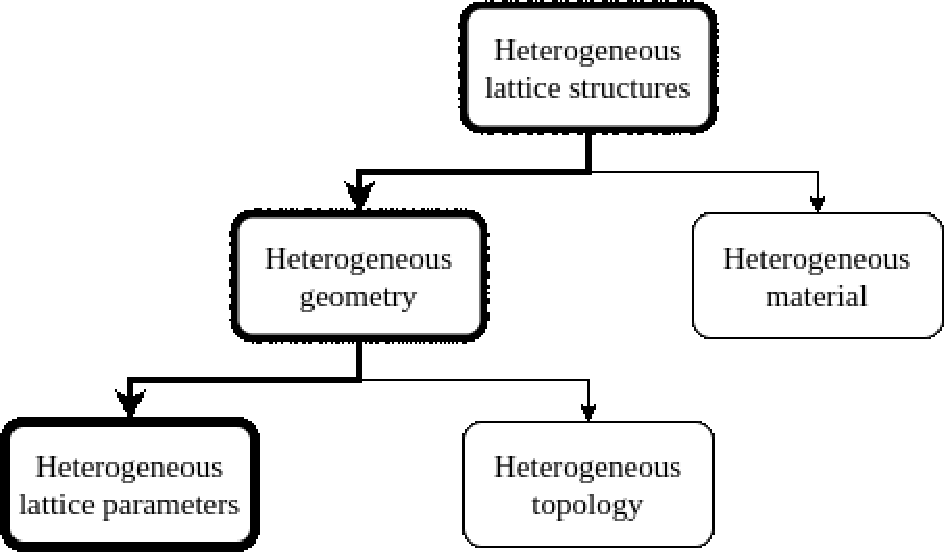}
    \caption{Various ways to parametrise
    heterogeneous lattice structures with
    the direction chosen for the proposed work
    encircled with bold lines}
    \label{fig:heterogeneity}
\end{figure}

The AM industry is currently in a situation where manufacturing
freedom exceeds design freedom~\parencite{Letov2021}.
The reason for this gap lies in significant challenges associated with
the geometric modelling of complex shapes.
Note that even though material heterogeneity is a significant
topic of interest in
CAD research~\parencite{Liu2021, Biswas2004}, the proposed work
considers only geometrical heterogeneity, such as the lattice structure
in Fig.~\ref{fig:hetero-example}.
Furthermore, at this research stage, it
is decided to focus on the parametrisation of lattice
parameters.
Figure~\ref{fig:heterogeneity} illustrates different
research directions on heterogeneous lattice structures in a diagram.
Lattice structures can be heterogeneous by their geometry and material.
The geometry can be made heterogeneous by varying lattice parameters
and topology.
The chosen direction is highlighted with bold borders and focuses
on heterogeneous lattice parameters.

This paper introduces a novel F-rep methodology
for the geometric modelling of heterogeneous lattice structures,
as well as its implementation in a software prototype.
The paper is organised as follows.
Section~\ref{sec:review} describes the concepts
related to the proposed work and reviews the related literature.
Section~\ref{sec:proposed-approach} explains the proposed approach
in geometric modelling terms.
Section~\ref{sec:implementation} covers the technical side of
implementing the proposed method in a software prototype.
The work is concluded in Section~\ref{sec:conclusions} with
analysing the results and performance of the proposed
method and its
implementation.

\section{Review of related literature}
\label{sec:review}

Before engaging in the development of the proposed approach,
an analysis and review
of related literature are performed.
Section~\ref{sec:representation} introduces surface and volumetric
modelling techniques that are common in CAD.
Section~\ref{sec:geometric-modeling} discusses the advancements in the
geometric modelling of heterogeneous lattice structures.
Section~\ref{sec:frep} describes the terminology
of functional representation of solid bodies used in this work.

\subsection{Geometric representation of solid models}
\label{sec:representation}
The majority of CAD tools utilise either the surface or
the volumetric modelling to represent the geometry of a solid body that
forms a CAD file.
These two approaches of the
geometric modelling are covered below in this subsection.

Any engineering software that needs to render a 3D geometric model, whether it is
for CAD, computer-aided manufacturing (CAM), or computer-aided engineering (CAE)
purposes has a geometric modelling kernel (GMK) at its core~\parencite{C3DLabs2020}.
The main goal of a GMK lies in building numerical models of required geometries
via mathematical methods and constraints.
Additional tools can be built on top of or alongside a GMK to provide additional
functionality.

There are plenty of literature review papers discussing the geometric
modelling approaches in
detail~\parencite{salomons1993review, kou2007heterogeneous}.
In this subsection, the key strategies are
summarised to justify the research gap
in the geometric modelling of heterogeneous lattice structures.

\subsubsection{Surface modelling approach}

The vast majority of the modelling techniques used in GMKs involve
the modelling of
surfaces that bound the solid model.
The most popular modelling techniques are boundary representation (B-rep)
and constructive solid geometry (CSG).
These techniques often utilise polygonal meshes, which operate
in the state of a trade-off between quality and performance.
It is not trivial to balance the trade-off when modelling
complex geometric
objects such as bio-inspired structures and heterogeneous lattice structures,
which often results in a high computation time and
decreased quality of models due to errors~\parencite{Letov2021,Cutanda2001}.

The CSG technique can be applied to define boundary conditions of the design
space of a lattice structure~\parencite{wang2021constructive}.
However, this research does not focus on defining the boundary conditions.
At the same time, the CSG technique is not well suited for
the geometric modelling of periodic structures~\parencite{loh2018overview}.

The CAD market is mainly dominated by the tools utilising B-rep.
B-rep allows setting limits that define the surface of the desired 3D model.
However, B-rep suffers from a lack of parametrisation and numerous operations
needed to achieve the modelling of even a homogeneous lattice~\parencite{Letov2021}.
The optimisation of the B-rep method introduced by hybrid
B-rep methods and optimisation
of boundary splines (B-splines) are still limited by the number of
operations
and parameters that are needed, even when the operations are
simplified~\parencite{Wang2014, Sasaki2017}.
Non-uniform rational basis splines (NURBS) and their extension are often used
to mitigate these issues and
the difficulties associated with the B-rep modelling by enabling
surface interpolation by trimming~\parencite{Rogers2001}.
However, since it is not trivial to obtain
a solution for a reverse problem, not
all surfaces can be trimmed to obtain the desired
shape~\parencite{Schmidt2012}.

\subsubsection{Volumetric modelling approach}

The volumetric modelling approach, contrarily
to the surface modelling approach, models not only
the surface $F(\mathbf{X}) = 0$ but the whole internal structure
$F(\mathbf{X}) \geq 0$ as well.
Modelling of the internal structure
enables better control over the geometric model since trimming and Boolean
operations are less computationally expensive
as there is no need to
generate a boundary surface~\parencite{Aremu2017}.
Such flexibility, however, often comes at the cost of
higher computation time to generate
volumetric models.
The advantages and disadvantages of voxel modelling have been reviewed in
literature extensively~\parencite{Liu2021, Letov2021}.

Voxels often have a cubic shape~\parencite{Strand2004}, which is then
smoothed to mitigate the effect of sharp corners by algorithms such as 
marching cubes~\parencite{Newman2006}.
Such representation commonly results in an inaccurate representation of
the model, with the only solution being using more voxels to ensure
a higher density of the building blocks of the model.
The increased amount of elements to compute
results in a much more time-consuming process due to the
computational complexity of voxel models being
$O(n^3)$~\parencite{Adalsteinsson1995}.
As a way to mitigate this complexity, a level-set method (LSM) is
used to decrease the dimensionality of the problem to $O(n^2)$,
which still can be highly time-consuming~\parencite{Adalsteinsson1995}.
Sparse voxel octrees also attempt to simplify the voxel rendering process
by combining large chunks of voxels, thus effectively reducing
their amount~\parencite{Laine2011}.

Voxels are not the only building blocks used in volumetric modelling.
The finite volume method (FVM) is another volumetric modelling method
inspired by the finite element method (FEM).
The main difference between FVM and FEM lies in using polyhedrons instead
of polygons as finite elements.
However, the FVM approach inherits all major disadvantages from FEM,
which prevents its practical use for geometrically complex
shapes~\parencite{Rom2011, Letov2021}.

\subsubsection{Hybrid modelling approaches}
Advantages of surface-based and volumetric geometric
modelling methods have inspired a search for hybrid
geometric modelling methods.
The AM process for the surface-based solids is
well-developed, and they provide significant ease of
manipulation with geometry. 
At the same time, voxel-based models enable computational
flexibility by allowing operations with 3D matrices. 
These advantages of surface-based and volumetric geometric
modelling methods have inspired a search for hybrid
geometric modelling
methods~\parencite{tang2019hybrid, wang2005hybrid}.

\subsection{Geometric modelling approaches for heterogeneous
lattice structures}
\label{sec:geometric-modeling}

Lattice structures have been a significant topic of
interest in the AM field,
as reviewed in Section~\ref{sec:introduction}.
This section examines geometric modelling methods applicable to heterogeneous
lattices and bio-inspired structures.
Note that the geometric modelling of bio-inspired structures
can prove helpful for
heterogeneous lattice structures since lattice structures are bio-inspired
themselves.
For example, the internal structure of bone, the Venus' flower basket, a bee
honeycomb and spider webs can be considered lattice structures.
These geometries evolved to provide lightweight yet durable
structures.

Naturally, as the AM industry expresses its interest in
the modelling of heterogeneous
lattice structures, several tools have emerged from the
research activities of academic institutions and CAD providers.

A common issue in the majority of lattice modelling tools
and CAD tools, in general, is that they are commonly supported by the Microsoft Windows
operating system (OS) only.
This limitation is reasonable as the majority of GMKs
that are behind all the mathematics and rendering have been developed on
and for Windows.
GMKs heavily rely on a graphic processing unit (GPU) 
to carry parallel computing within, and
Windows has time-proven proprietary drivers for GPUs.
However, Apple macOS recently got reinforced
by the  M1 central processing unit (CPU), which can handle parallel computing
similarly to a GPU~\parencite{cryptoeprint:2021:986}.
At the same time, Linux allows far greater potential in
customising the system to specific needs,
and proprietary GPU drivers can be installed with ease while
the OS performance remains low compared to Windows.
Both macOS and Linux are based on the Unix OS family which
makes the process of developing software
tools that are cross-platform between them easier.
With this in mind, it can become more common to see
geometric modelling tools being cross-platform in the future.

This section covers the application of geometric
modelling approaches for heterogeneous lattice
structures.

\subsubsection{Surface modelling approach}

Autodesk Netfabb~\parencite{netfabb}
is one of the tools that allow the modelling of lattice
structures.
To model a lattice structure with Autodesk Netfabb, one must
provide a design space or choose one of the standard types
of design spaces.
Next, a topology must be chosen from a list of supported topologies.
This list is substantial and includes topologies based on beams and triply
periodic minimal surfaces (TPMS), but it cannot be extended by the user.
There is an option to apply a linear gradient field to the design space so that
the thickness of the lattice structure would be varying
according to that field.
Still, this gradient does not allow changing of other lattice parameters
and does not support variation of parameters beyond linear.

Sulis Lattice~\parencite{sulis}
allows importing of a CAD model and the use of its regions
as a design
space for the lattice generation.
A list of topologies to choose from is present, as well as the ability
to add a linear gradient distribution of the lattice thickness.
As of now, there is no support for non-linear thickness
distribution and the variation of parameters
other than beam or surface thickness.

\begin{figure}
    \centering
    \begin{subfigure}[b]{\subfigw}
            \centering
            \includegraphics[width=\textwidth]{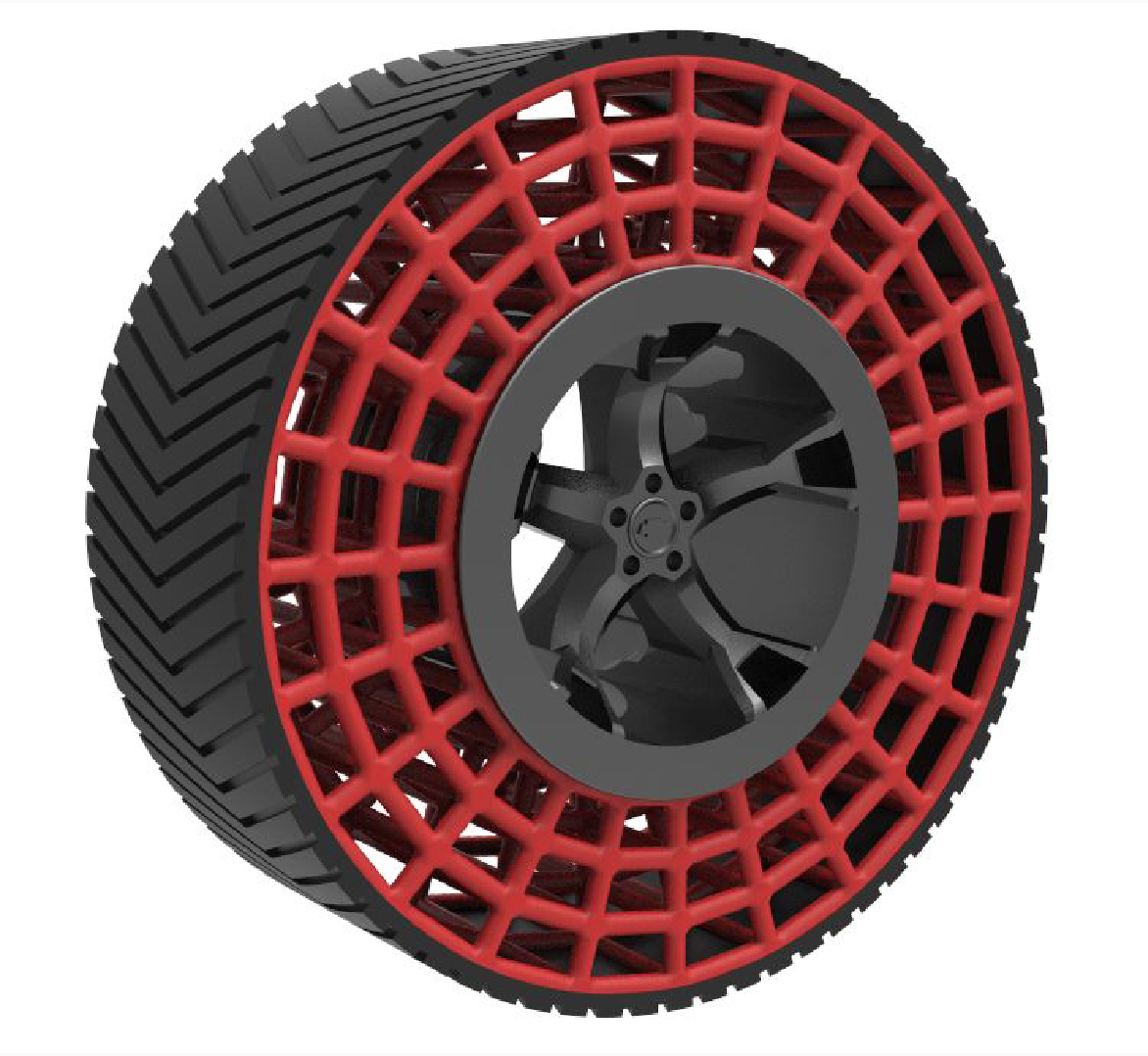}
            \caption{}
            \label{fig:tire-grid}
    \end{subfigure}
    \hfill
    \begin{subfigure}[b]{\subfigw}
            \centering
            \includegraphics[width=\textwidth]{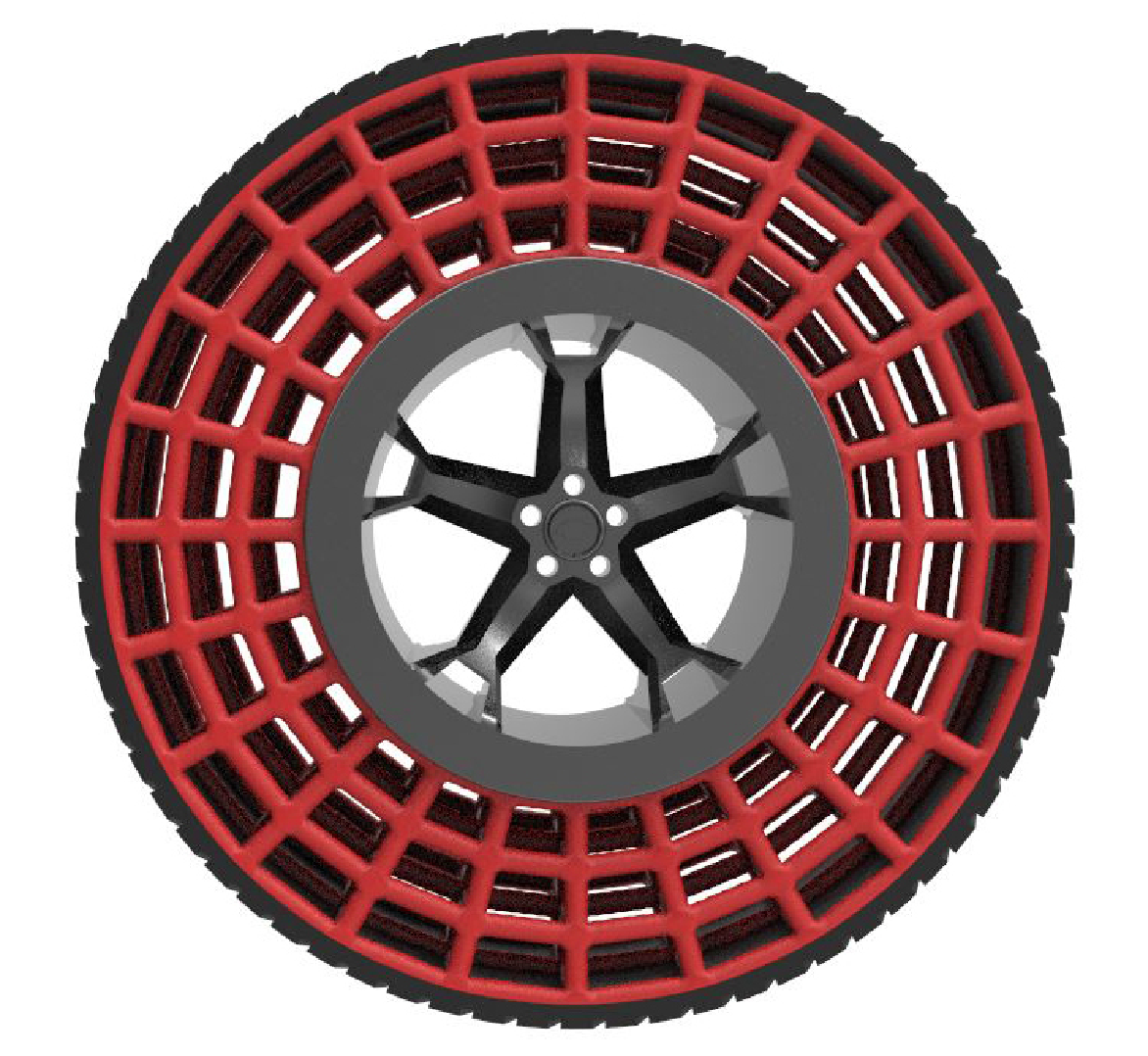}
            \caption{}
            \label{fig:tire-grid-prof}
    \end{subfigure}

    \caption{(a) An isometric and (b) a profile view on a conformal
    lattice tire design modelled with Intralattice~\parencite{Kurtz2015}}
    \label{fig:intralattice-tire}
\end{figure}

Rhinoceros 3D~\parencite{rhinoceros}
is another tool that is highly adaptable for the modelling
of heterogeneous lattice structures.
Rhinoceros 3D
is a parametric CAD that allows tuning geometric
parameters and scripting of the design with embedded Python support.
Grasshopper 3D~\parencite{grasshopper}
within Rhinoceros 3D allows visual scripting of
parametric models.
Moreover, Grasshopper 3D supports custom plugins that can aid with 
lattice design.
One of the research efforts preceding this work resulted
in the development of Intralattice from this research
team~\parencite{Kurtz2015}.
Intralattice can model lattice structures with
custom topologies.
However, the heterogeneity of the lattice structures modelled with
Intralattice is limited to conformal lattice
structures -- pseudoperiodic lattice structures that strive to fit a
custom design space.
For example, the tire design illustrated in
Fig.~\ref{fig:intralattice-tire} has the design space filled with a
conformal lattice which can be considered
homogeneous in cylindrical coordinates.
Heterogeneity of other geometric lattice parameters is not supported.
Crystallon~\parencite{FEQUALSFLLC2019}
is another plugin for Rhinoceros 3D that allows conformal
lattice modelling with a topology chosen from a list of available
ones~\parencite{Letov2021}.
Similarly to Intralattice, there is no support for
the modelling of heterogeneous lattice structures in Crystallon.
Interestingly, a use case of applying both Intralattice and 
Crystallon for the same lattice generation project is reported
in the literature~\parencite{Garcia-Dominguez2020}.
The reason for this symbiosis lies in the relative ease of defining
topologies with Intralattice, while Crystallon was able to generate
nodes better.
As both Intralattice and Crystallon are not capable of modelling
heterogeneous lattice structures effectively, another plugin for
Grasshopper 3D -- Dendro~\parencite{dendro2018}
-- was developed to bridge this gap.
Dendro utilises voxel modelling techniques provided by OpenVDB to model
linearly heterogeneous lattice structures.
Nevertheless, non-linear heterogeneity lies beyond reach for
the existing plugins for Grasshopper 3D.

Subdivision surfaces have a strong potential in the
heterogeneous lattice structure modelling.
Originally, subdivision surfaces were used in computer graphics algorithms
that produce smooth B-rep
surfaces by subdividing the initial rough mesh into a denser interpolated
mesh~\parencite{Catmull1978}.
Subdivision surfaces allow setting several points that define a surface
to interpolate the whole surface.
This interpolation has been successfully applied for the geometric
modelling of beam-based structures~\parencite{savio2018geometric} and 
TPMS-based structures~\parencite{Savio2019}.
However, this approach has been primarily manual so far,
and the geometric parameters
need to be specified individually for each unit cell.
The rising amount of manual operations
means that the modelling complexity rises proportionally to the number
of unit cells that can reach hundreds.
Weaverbird~\parencite{weaverbird} -- a plugin for
Grasshopper that allows converting
the surface of a solid into a lattice -- is  based on
subdivision surfaces.

Subdivision surfaces can be applied not only for
the modelling of a single B-rep surface
but to an individual face of a surface.
This approach is known as the T-splines modelling and
allows having denser mesh in
the most critical regions of the model~\parencite{sederberg2004t}.
T-splines have found their application in the modelling
of periodic geometric
models~\parencite{wang2021characterisation}
and conformal lattice structures~\parencite{xiao2019interrogation}.
These successful applications motivate
further investigation of applying T-splines
to the geometric modelling of heterogeneous lattice structures.

\subsubsection{Volumetric modelling approach}

Voxel-based modelling is gaining popularity for its applications to
heterogeneous lattice structure modelling.
For conventional manufacturing, choosing a correct voxel size can be
ambiguous, especially with solid models with a high degree
of curvatures.
However, for AM, there is no need for the voxel size to exceed the
tolerance of a 3D printer since the higher voxel resolution would not be
manufacturable~\parencite{Letov2021}.
Assignment of material in the voxelised model of a multi-material
lattice structure is simple since every voxel can be assigned different
materials~\parencite{liu2022heterogeneous}.

Another way to obtain volumetric CAD models of heterogeneous
lattice structures
and other bio-inspired structures is the export of geometry
from computer tomography (CT) scans.
Some of the tools capable of retrieving these models include
Materialise Mimics~\parencite{mimics}
and Dragonfly~\parencite{dragonfly}.
Materialise Mimics segments an imported medical image which
can be converted to a solid model and filled with a lattice structure
by the Materialise 3-matic tool.
Similarly, the output solid models in Dragonfly can be subject to
various manipulations with the data.
For example, the porous bone structure can be viewed as a graph
with all the nodes and links extracted~\parencite{Reznikov2020}.
Such tools provide a different perspective on the knowledge of 
biological structures.
These tools are often used for AM of prosthetic parts.
Notably, biological structures are often heterogeneous, but the current
tools often lack high-tuned control over geometric parameters
of heterogeneous lattice structures~\parencite{Letov2020}.

\subsubsection{Hybrid modelling approaches}

Implicit modelling approaches are extremely powerful when applied in the
geometric modelling of heterogeneous lattice structures.
The explicit modelling techniques would require each beam or surface region
of a lattice structure to be modelled explicitly.
Meanwhile, the implicit modelling techniques provide a set of high-level functions
that simplify the process of the definition of complex geometry by packing lower-level
functions together into one~\parencite{nguyen2021implicit}.

nTopology~\parencite{ntopology} is a heterogeneous lattice
modelling tool that is gaining popularity.
Similarly, to Autodesk Netfabb, design space and a topology
need to be set, usually by importing a CAD file with distinguishable
lattice features.
However, the topologies in nTopology are defined as skeletal graphs,
which allows their thickening to obtain a solid model of the desired
lattice.
This approach is based on voxelising the space
neighbouring the skeletal graph
and adding or deducting extra layers of voxels to increase or decrease the thickness.
In addition to the linear variation of the thickness as in Autodesk
Netfabb,
nTopology introduces topology optimisation to control the thickness based
on estimated stresses.
Topology optimisation can be considered an implicit
geometric modelling method, as the user input is limited, and
there is little control over geometric parameters.
Instead, the geometric parameters depend on the result of a CAE
simulation.
nTopology relies on a graphics processing unit (GPU) for the
acceleration of the solid model preview. The output itself can be
a mesh exported from a voxel grid.
nTopology supports export to the stereolithography (STL) file format,
which is encoded with the American Standard Code for Information Interchange
(ASCII) and is the most popular
3D CAD file format for AM, and to
the STEP file defined by the ISO 10303-21
standard~\parencite{ISO10303} which is among the few CAD file formats
suitable for CAE simulation purposes~\parencite{hamri2010software}.

Randomised lattice structures such as the Voronoi scaffold illustrated in
Fig.~\ref{fig:voronoi} often require specialised approaches.
For example, consider a Voronoi scaffold that does not have any unit cell in common sense.
Designing a Voronoi scaffold requires setting general geometric parameters of the lattice
rather than geometric parameters of each unit cell.
These parameters include the number of Voronoi seeds and the size of
pores~\parencite{Fantini2016}.
Such randomised structures often appear in nature.
At the same time, beam-based lattice structures are not common in
nature, and an advanced surface
modelling approach is usually required to mimic
bio-inspired geometry appropriately.
It is possible, for example, to apply randomised geometric modelling
approaches based
on TPMS to mimic bone tissue~\parencite{Shi2018}.

Some of the existing tools support implicit \hl{modelling} to a certain extent.
Implicit \hl{modelling} enables the definition of geometry based on implicit
mathematical functions.
Implicit functions extend the design freedom of a geometric \hl{modelling}
tool by, for example, \hl{minimising} the user input
and is the only practical way to model TPMS
structures.
In particular, topology \hl{optimisation} supported by
nTopology \hl{is characterised as implicit modelling}.
Topology \hl{optimisation} adjusts the geometric parameters of each unit
cell of a lattice according to a field function.
This field function is commonly obtained through a CAE simulation.
Topology \hl{optimisation} is currently limited by \hl{optimising} the
thickness of a lattice which in some cases is not a sole parameter of
a topology.
The truncated cube topology, for example, requires the truncation parameter
to be fully defined.
Moreover, the design freedom is limited by the results of the
\hl{optimisation} process.
Other adjustments have to be introduced in the resulting solid model
manually, often to each \hl{individual} relevant unit cell.
This work mainly focuses on geometric \hl{modelling} methods
not based on \hl{optimisation}.

\hl{MATLAB}~\parencite{matlab}\hl{ is another powerful tool that finds
applications in mathematical and geometric simulations.
Moreover, MATLAB is extendable by add-ons.
For example, MSLattice}~\parencite{al2021mslattice}
\hl{allows the modelling of various TPMS-based lattice structures.
It also supports the modelling of a transition between two different topologies.
Another example is FLatt Pack}~\parencite{maskery2022flatt}
\hl{which allows their modelling
of simple TPMS-based homogeneous lattice structures and even their export
to the STL file format.
Both tools support the modelling of conformal lattice structures in cylindrical
and spherical coordinates.
FLatt Pack also supports a homogeneous lattice infill within
an imported STL file.
Both tools do not support the STEP file format export as of now.
Both tools are also limited to the modelling of TPMS-based lattice structures
only.
However, FLatt Pack considers the BCC topology as an extreme case of
a gyroid topology.
FLatt Pack and MSLattice define the TPMS topologies by their corresponding
implicit functions.
FLatt Pack voxelises the unit cell design space according to that implicit
function and can approximate the triangular mesh for export to STL.}

\subsection{F-rep concepts and their application in geometric
\hl{modelling}}
\label{sec:frep}

Function representation (F-rep) methods allow \hl{the modelling}
of a geometric object as a
set of mathematical functions~\parencite{Pasko1995}.
In F-rep, a solid is considered to be defined by its defining real-valued function
$F$ as follows
\begin{equation}
    F(\mathbf{X}) \geq 0,
\label{eq:frep}
\end{equation}
where $\mathbf{X} = (x, y, z) \subset \mathbb{R}^3$ is the design space, such
that $F(\mathbf{X}) \geq 0$ is the solid itself with $F(\mathbf{X}) = 0$ being
the surface of the solid, and $F(\mathbf{X}) < 0$ is the rest of the design
space~\parencite{Pasko1995}.
Note that the design space $X$ is not limited by existence in the 3D Euclidean
space $E^3$, as printing of non-Euclidean geometry is a topic of interest in
AM nowadays~\parencite{PhysRevLett, Mensch2021}.

Even though F-rep existed before the rise of AM, using real functions for
representing solid models demonstrated advantages in providing higher design
freedom compared to the conventional Boolean methods of
\hl{modelling}~\parencite{Shapiro1994}.
Modern F-rep methods also allow the \hl{modelling} of both explicit and implicit functions.
In this work, a function $F: \mathbb{R}^n \to \mathbb{R}$ is called explicit
if $F$ is defined by an expression, i.e. by a relation solved for one of its
independent variables, for example, $F(x):=x^2+2x+1$.
A function $F: \mathbb{R}^n \to \mathbb{R}$ is called implicit if $F$ is defined
by a relation not solved for one of its independent variables, for example,
$x^3F^3(x)=F(x)+2x$.

\begin{figure}
    \centering
    \includegraphics[width=0.7\textwidth]{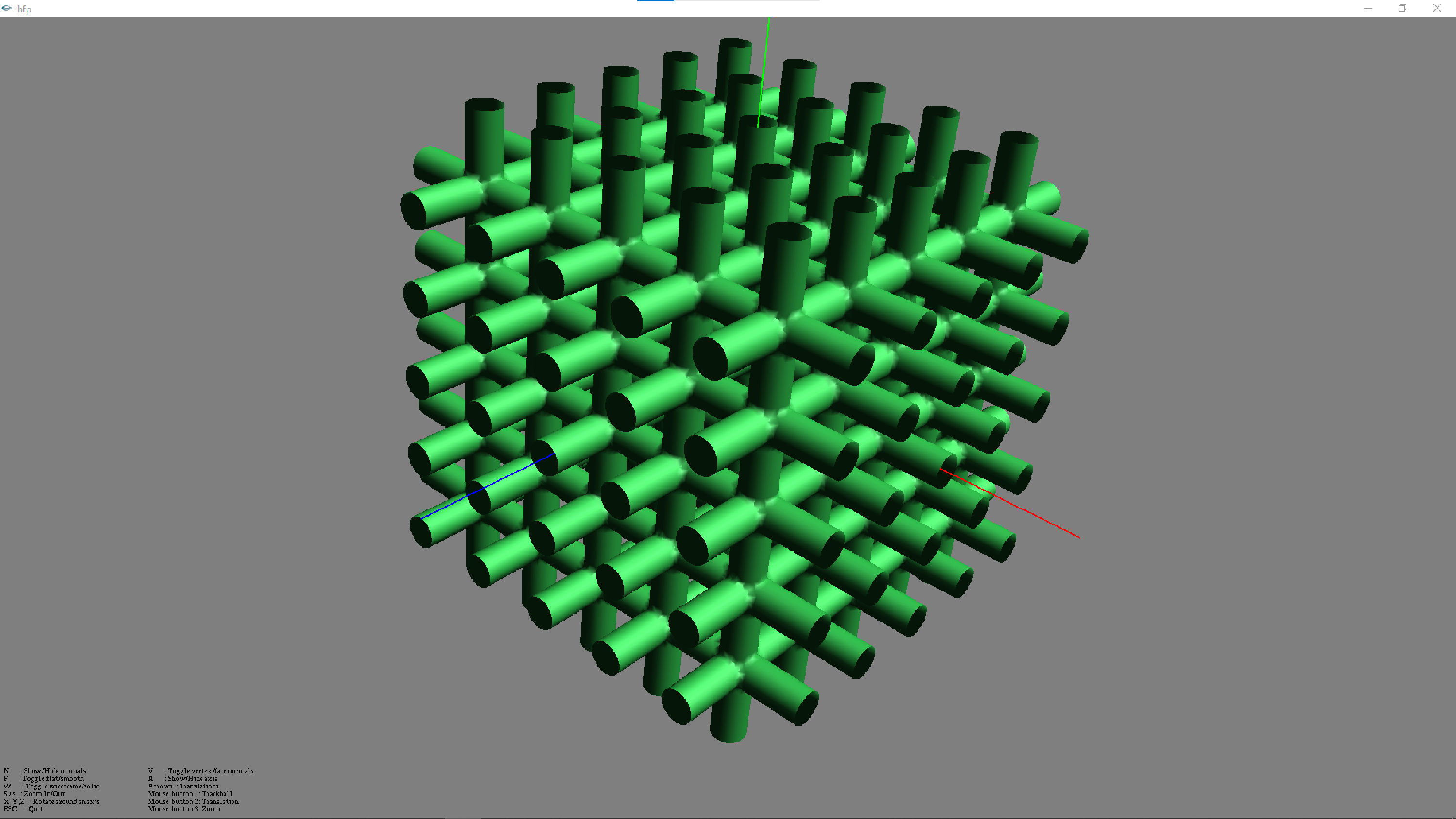}
    \caption{A homogeneous lattice structure with the simple cubic topology generated with HyperFun}
    \label{fig:lattice-hyperfun}
\end{figure}

Comparing $F(\mathbf{X})$ with $0$ in F-rep allows \hl{the modelling}
of implicit surfaces
bounding the solid body without calculations needed to convert them into explicit
surfaces.
For example, a solid torus $\mathbb{T}_S$ implicitly defined by
\begin{equation}
  \mathbb{T}_S(x,y,z): -F(\mathbf{X}) = \left(\sqrt{x^2 + y^2} - R\right)^2 + z^2 - r^2 \leq 0,
\label{eq:torus-implicit}
\end{equation}
where $x,y,z$ are the Cartesian coordinates, $R$ is the radius from the
\hl{centre}
of the hole to the \hl{centre} of the torus, and $r$ is the radius of the tube.
The torus $\mathbb{T}$ can be explicitly defined by its standard form derived
from solving $F(\mathbf{X})=0$ for $z$:
\begin{equation}
  \mathbb{T}(x,y,z): z = \pm\sqrt{r^2-R^2+2R\sqrt{x^2+y^2}-x^2-y^2}.
\label{eq:torus-explicit}
\end{equation}
Equation~\ref{eq:torus-explicit} requires additional calculations to
derive it from the standard form in Equation~\ref{eq:torus-implicit}.
While solving $F(\mathbf{X})=0$ for the torus $\mathbb{T}$ results in
Equation~\ref{eq:torus-explicit}, solving $F(\mathbf{X})
\geq 0$ for the solid torus $\mathbb{T}_S$ requires more conditions that need
to be taken into account.
Moreover, there are cases when an explicit form of the function cannot be
achieved, e.g. functions defined by $x^2y^2=(x+y)^3-\sqrt{xy}$,
$x\sqrt{cos{(xy)}}=e^y$, etc.
Thus, the ability to use implicit functions as an input is a significant
advantage of F-rep\hl{,} which \hl{amplifies further because}
lattice structures can be defined as a set of functions.
For example, HyperFun -- an F-rep programming language -- can model a simple
homogeneous lattice structure illustrated in Fig.~\ref{fig:lattice-hyperfun} in a simple loop~\parencite{pasko1999hyperfun}.
Note that this representation is simplistic as nodes are not modelled\hl{,} and
the lattice structure is homogeneous.
Also, note that HyperFun is not publicly available now and is now available from
Uform~\parencite{hyperfun_2020}.
\hl{The ability to render periodic structures and mathematically
well-defined Boolean operations between functions allowed F-rep
to find its application for the modelling of heterogeneous lattice
structures
}\parencite{alkebsi2021design, wang2022computational}.

Initially, implicit surfaces used in F-rep had limited support of $R$-functions,
sweeping and other operations \hl{that are} common for conventional CAD systems.
To mitigate this limited support, F-rep methods use non-trivial
solutions to this issue\hl{,} such as moving solids~\parencite{Sourin1996}.
The $R$-function support drastically \hl{increases} the flexibility of geometric
\hl{modelling} by allowing \hl{a} complete definition of
geometry with the terms of real
analysis~\parencite{Shapiro2007}.
Naturally, such a viable tool \hl{is} now found supported by most of the
F-rep geometric \hl{modelling} approaches.

Still, there are certain disadvantages \hl{to} using \hl{the F-rep modelling}.
Defining a geometrically complex structure such as a heterogeneous lattice structure
is complicated by the necessity of defining rules by which lattice parameters
or topology vary throughout the structure in a set of mathematical functions.
The review of related literature identified that such a process can be
\hl{arduous} for an engineering designer and that there is no tool so far that would
simplify this design process significantly enough, while it is clear that AM could
benefit from such a tool~\parencite{gandhi2015designing, Liu2017}.
Moreover, the function by which a geometrical shape is formed is not defined
clearly in some cases\hl{,} such as, for example, in bio-inspired design~\parencite{Letov2021}.

\section{The proposed F-rep approach}
\label{sec:proposed-approach}
As described in Section~\ref{sec:frep},
F-rep \hl{modelling} methods require
real mathematical functions to describe the geometry of a solid body.
\hl{Such definition of geometry}
provides significant design freedom since any shape can be defined by a
mathematical function or an interpolation of one~\parencite{Savchenko1995, Letov2021}.
However, \hl{utilising} such design freedom can prove challenging due to the process
of defining the functions being difficult and tedious for an engineering designer.
This challenge is usually solved by \hl{utilising} simpler interpolations for the
functions that define \hl{the} geometry~\parencite{Savchenko1995, YeungYam2006}.
Such approximations, however, result in a solid model that is not of the original
design.
An approximated model often implies a geometry that does not reach all the
goals set for the product before the conceptual design
process~\parencite{Hsu2000, Letov2021}.

\begin{figure}[h]
    \centering
    \includegraphics[width=0.45\textwidth]{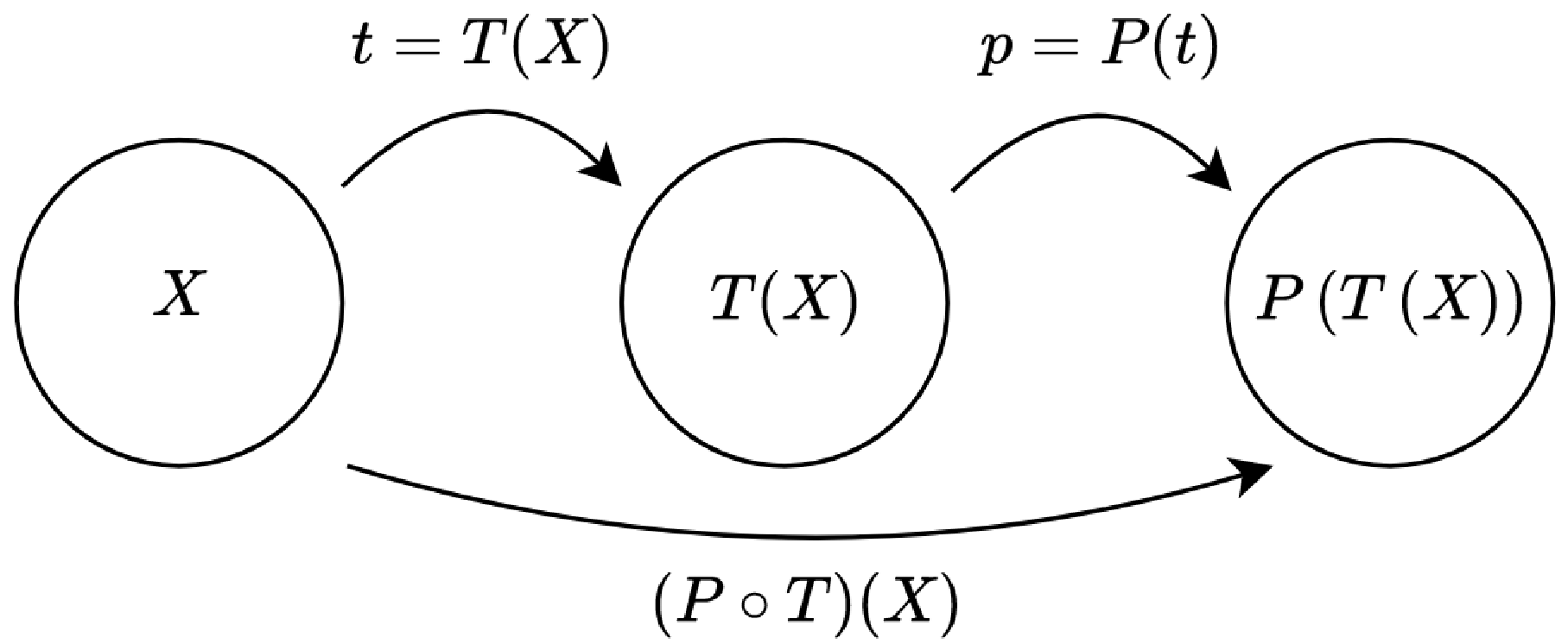}
    \caption{The sequential mapping of topology $T$ and parameters $P$ to form a single
    heterogeneous lattice structure}
    \label{fig:mapping}
\end{figure}

In this research, instead of defining a complex set of functions that
define the geometry, the focus was placed
on providing tools to define these functions
\hl{more straightforwardly}.
In the proposed method, a single function approach is implemented with the
function depending on parameters necessary for \hl{the modelling} of a
heterogeneous lattice structure.
Moreover, parameters themselves are proposed to be controlled by functions.
First, such an approach supports the design freedom by allowing
the variation of lattice parameters such as the beam or node diameter,
surface thickness, etc.
Secondly, this approach can potentially allow the \hl{modelling} of
hierarchical lattice structures since lattice defining function
can be nested inside a higher tier function.
It is proposed to expand the conventional F-rep approach
to be better suited for \hl{the modelling} of heterogeneous lattice structures.
Instead of a single function $F(\mathbf{X})$ that defines the solid model,
it is proposed to split this function into two: one function 
that defines the shape of a unit cell, and another one
that defines \hl{the} geometric parameters of that topology.
So, this special case of the conventional F-rep model defined
by Equation
\ref{eq:frep} is defined \hl{in this work} as
\begin{equation}
    F(\mathbf{X}) = (P \circ T)(\mathbf{X}) \geq 0,
\label{eq:composition}
\end{equation}
where $T$ defines the topology of the lattice and $P$ defines
\hl{the} parameters of the topology.
Figure~\ref{fig:mapping} illustrates the heterogeneous lattice mapping process.
Note that the order of the composition in Equation~\ref{eq:composition} matters,
as the parameters are different for different topologies.

This approach allows \hl{the modelling} of highly complex solids.
In this research, the complexity of a solid body $\mathbb{S}$
\hl{that is} defined by
\begin{equation}
    \mathbb{S} := \{ \mathbf{X}| F(\mathbf{X}) \geq 0\}
\end{equation}
is considered to correlate with the genus of its bounding surface
$\partial \mathbb{S}$ defined by
\begin{equation}
  \partial\mathbb{S} := \overline{\mathbb{S}} \cap \overline{(\mathbf{X}-\mathbb{S})}= \{ \mathbf{X}| F(\mathbf{X})=0\},
\end{equation}
which is essentially an oriented 2-manifold $M^2_g$ of a finite genus $g$.
Here $\overline{\mathbb{S}}$ is the closure of $\mathbb{S}$ and
$\overline{(\mathbf{X}-\mathbb{S})}$ is the closure of the compliment of
$\mathbb{S}$.
The complexity of a lattice structure is often estimated by its genus $g$ since
it correlates with the amount of non-trivial curvatures in the corresponding solid model~\parencite{Feng2018, Letov2020}.
For example, a single simple cubic unit cell $\mathbb{U}_\text{cubic}$ of a lattice 
structure in Fig.~\ref{fig:lattice-hyperfun} has genus $g=5$ since it is
homeomorphic to $M^2_5$ or 5-torus $\mathbb{T}^5$:
\begin{equation}
    \mathbb{U}_\text{cubic} \cong \mathbb{T}^5 = S^1 \times S^1 \times S^1 \times S^1 \times S^1,
\end{equation}
where $S^1$ is a circle. 
The complexity of a solid torus $\mathbb{T}_S$ -- the most simplistic solid body
with a single hole -- is proportional to $g=1$ of the torus $\mathbb{T}\cong M^2_1$.
Gyroid lattice structures can be of different varieties with the genus of at least
$g=5$~\parencite{Gozdz1996}.

The proposed framework allows union operations between solids, such that
$\mathcal{S} = \bigcup_{i=1}^{n} \mathcal{S}{i}$,
where $n$ is the total number of \hl{solids} and $\mathcal{S}{i}$ is the $i$-th solid.
In the case of trivial union, the transition between topologies can become abrupt 
and the border between topologies might not be smooth unless there is a perfect match of nodes.
The topology transition in lattice structures already is a topic of interest
in heterogeneous lattice research and has been implemented previously
in several works~\parencite{Yang2015c}.
\hlcyan{In particular, the transition between two surface-based topologies
invokes the most challenges}~\parencite{kim20203d}.
The proposed F-rep framework at its current stage proposes
control of \hl{the} topology parameters $P(\mathbf{X})$ only.

The implementation of the proposed framework is detailed in
Section~\ref{sec:implementation}
which clarifies the technical approaches used to develop a software prototype of the
framework.

The rest of this section discusses the proposed F-rep approach in detail.
Section~\ref{sec:topologies} introduces the proposed geometric description $T$
to represent lattice topologies.
Section~\ref{sec:parameter-variation} describes the proposed approach for the
variation of geometric parameters $P$ in a non-linear manner.

\subsection{Functional definition of lattice topologies}
\label{sec:topologies}

The proposed framework is developed in such a way that it
supports both beam-based and surface-based topologies.
The topology is proposed to be defined by its skeleton
which is defined by function $T$.
Subsections~\ref{sec:beams-framework} and~\ref{sec:tpms-framework} provide
a description of how the proposed approach can be used for \hl{the modelling} of
beam-based and TPMS-based topologies, respectively.

\subsubsection{Beam-based topologies}
\label{sec:beams-framework}
Since the methodology described in this work is proposed to be based
on F-rep, functions need to be defined for the common \hl{topologies}.
\hl{A skeleton of a beam-based topology} can be defined by a set of
lines that are defined in $x,y,z\in[0,u]$, where $u$ is the size of a
unit cell.

\begin{figure}
    \centering
    \includegraphics[width=0.45\textwidth]{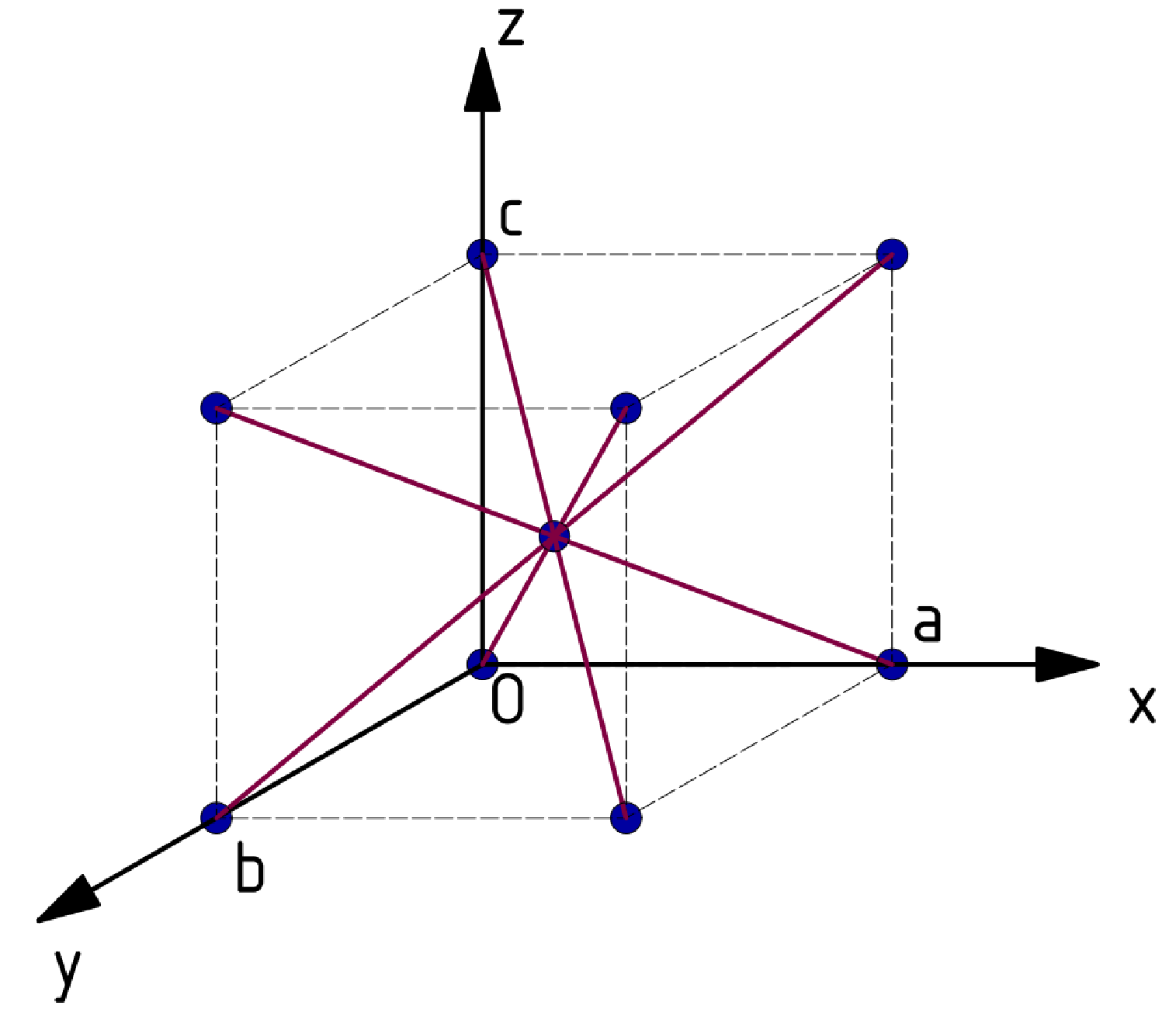}
    \caption{A BCC unit cell described by Equation~\ref{eq:bcc}}
    \label{fig:bcc-sketch}
\end{figure}

As an example of how a beam-based topology can be defined, consider
a \hl{body-centred} cubic topology (BCC) sketched in Fig.~\ref{fig:bcc-sketch}
with 8 nodes in every vertex of a cube and one more node in the \hl{centre}
of the cube.
As covered in a preceding work~\parencite{Letov2021}, the skeletal frame for this topology
can be defined as follows:
\begin{equation}
    T(\mathbf{X}):
    \left[ 
      \begin{gathered} 
        \frac{x}{a}=\frac{y}{b}=\frac{z}{c}, \\ 
        \frac{x-a}{-a}=\frac{y}{b}=\frac{z}{c}, \\
        \frac{x}{a}=\frac{y}{b}=\frac{z-c}{-c}, \\ 
        \frac{x-a}{-a}=\frac{y}{b}=\frac{z-c}{-c} \\ 
      \end{gathered} 
    \right.
    \text{for } x\in[0,a],y\in[0,b],z\in[0,c],
    \label{eq:bcc}
\end{equation}
where $a$, $b$, and $c$ are sides of a cuboid unit cell.
In case of a cubic unit cell, $a=b=c$.
After defining the skeletal frame, a cross-section with varying 
parameters can be assigned to the frame.
Cross-section itself can be defined by function $c(\mathbf{X}_c)$, where
$\mathbf{X}_c\subset\mathbb{R}^2$ is the coordinate space local to the
cross-section plane.
For example:
\begin{itemize}
\item A cylindrical beam can be defined by a cross-section
which can be defined as
\begin{equation}
    c(x_c,y_c): x_c^2+y_c^2=R_c^2,
    \label{eq:circle}
\end{equation}
where $x_c, y_c\in\mathbb{R}^2$ are 
Cartesian coordinates local to the cross-section plane and $R_c$
is the radius of the circular cross-section.
\item A beam with a square cross-section can be defined as
\begin{equation}
    c(x_c,y_c): |x_c+y_c|+|x_c-y_c|=\tau,
    \label{eq:square}
\end{equation}
where $x_c, y_c\in\mathbb{R}^2$ are 
Cartesian coordinates local to the cross-section plane and $\tau$
is the side of the square.
\item A beam with a cross-section that has a shape of a square with rounded
corners can be defined as
\begin{equation}
    \begin{gathered}
    c(x_c,y_c):\max{\left(|x|-\frac{\tau}{2}+\rho, 0\right)}^2\\
    + \max{\left(|y|-\frac{\tau}{2}+\rho, 0\right)}^2 = \rho^2,
    \end{gathered}
    \label{eq:rounded-square}
\end{equation}
where $x_c, y_c\in\mathbb{R}^2$ are 
Cartesian coordinates local to the cross-section plane, $\tau$
is the side of the rounded square, and $\rho$ is its fillet radius.
Note that Equation~\ref{eq:rounded-square} converges to Equation~\ref{eq:circle}
with $\rho\to\rho_{max}=\tau/2$ and
to Equation~\ref{eq:square}
with $\rho\to\rho_{min}=0$.

\end{itemize}

F-rep in the proposed work is proposed to allow control
over these parameters, thus allowing change of topology parameters
throughout the whole structure.
For example, the diameter of beams within a lattice with the BCC
topology can be controlled by function $P$.

This way of defining skeletal graphs is based on F-rep which makes it
different from the voxel-based method used in nTopology.
F-rep allows a more straightforward way of defining the thickness of topology
by defining a variable within the function ($R$ for the circle $c$ in
the example above).
Furthermore, as described further, this approach
allows variation of parameters other than thickness, thus further extending
the design freedom.

Various lattice topologies were defined according to the proposed approach.
\hl{Appendix}~\ref{sec:tables} lists the topologies and the functions
that define them \hl{following} the proposed approach.
Table~\ref{tbl:topologies} lists topologies inspired by the cubic crystal system
in crystallography.
These topologies are known for their ability to reinforce the lattice structure
they are applied in specific directions~\parencite{maskery2017investigation}.

There are other topologies that are not
inspired by the cubic crystal system but which are beam-based
such as diamond, rhombicuboctahedron, and truncated cube.
Their F-rep functions are listed in
Tables~\ref{tbl:diamond}--\ref{tbl:tcube} in Appendix~\ref{sec:tables}.
Note that the rhombicuboctahedron and truncated
cube topologies require an additional truncation
parameter $t\in[0, 0.5u]$ which sets the size of
truncation.

\subsubsection{TPMS-based topologies}
\label{sec:tpms-framework}
\begin{figure}[h]
    \centering
    \includegraphics[width=0.4\textwidth]{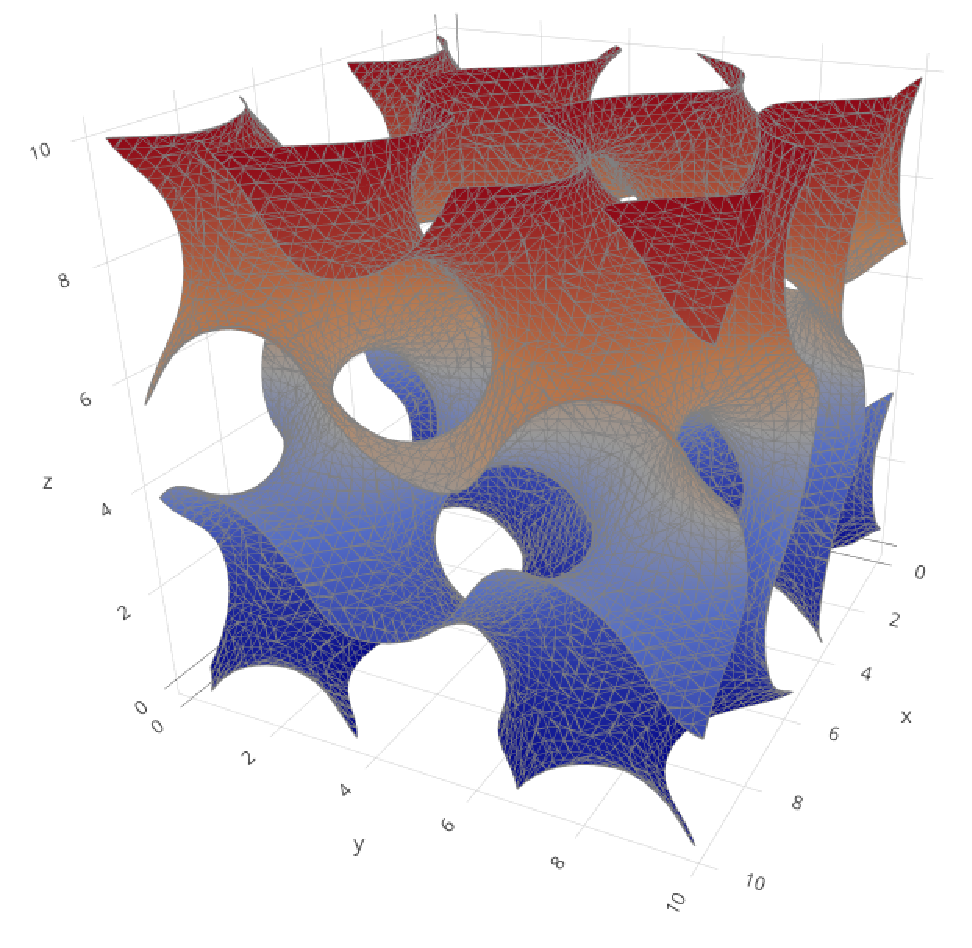}
    \caption{A gyroid surface described by Equation~\ref{eq:gyroid}}
    \label{fig:gyroid-skeleton}
\end{figure}

The same approach is proposed to be applied to the topologies based
on TPMS.
For TPMS-based lattices, the skeletal frame is equivalent
to the TPMS itself described by a mathematical equation.
Since the approximations of these equations are well-known,
there is even more reason to rely on F-rep for \hl{the modelling} of TPMS-based
lattices.
For example, a gyroid lattice illustrated in Fig.~\ref{fig:gyroid-skeleton}
is defined by its equation as follows:
\begin{equation}
    \sin(x)\cos(y) + \sin(y)\cos(z) + \sin(z)\cos(x) = 0.
    \label{eq:gyroid}
\end{equation}

Unlike the beam-based topologies, TPMS-based topologies do not require
its cross-section to be defined as a function $c$.
Instead, only the thickness of a solid body is required to be defined.

Table~\ref{tbl:tpms-topologies} covers 
the topology defining functions $T$ for the TPMS-based topologies.

This approach allows the proposed framework to not be limited to beam-based topologies,
as F-rep allows \hl{the modelling} of everything that can be defined with functions that
define the geometry.

\subsection{Functional variation of geometric parameters}
\label{sec:parameter-variation}

As mentioned in Section~\ref{sec:review}, most existing approaches for \hl{the
modelling} of heterogeneous lattice structures permit control of the lattice thickness.
This control is \hl{primarily} gradient-like\hl{,} with the linear distribution of the thickness
in a specified direction.
The only exception among the reviewed methods is the topology
\hl{optimisation}.
\hlcyan{While topology optimisation allows the generation of geometrically highly
complex structures}~\parencite{liu2021stress}, \hlcyan{it has limited control
from the user.
It was decided to develop an implicit modelling method that is not based on topology
optimisation.}

The proposed approach allows variation of various geometric parameters in
different directions and is not limited to the linear distribution of the lattice
parameters.
This \hl{variation} is enabled by the introduction of the
function $P$ \hl{which controls} the parameters.
For example, $P(\mathbf{X}): t(z)$, where $t: \mathbb{I}\rightarrow\mathbb{R^+}$
is the thickness of the lattice\hl{,}
and $\mathbb{I}$ is a unit interval $[0,1]\subset\mathbb{R}$.
This approach allows setting thickness $t$ as a distribution defined by any
mathematical function.
Here, \hl{the} thickness can be either the beam thickness or the surface thickness,
depending on function $T$ that \hl{describes} the topology.
Several use-cases of varying $P(\mathbf{X}): t(z)$ are presented
in Figures 12-15 in
Section~\ref{sec:implementation}.

Geometric parameters other than lattice thickness can be controlled with
the proposed approach.
Such parameters include, for example, the radius $\rho$ of the fillet of
the square beam with the rounded corners described in Equation
\ref{eq:rounded-square}.
The truncation parameter in the rhombicuboctahedron and the truncated cube topologies
is another example of a geometric parameter other
than the lattice thickness.
Both topologies converge to simple cubic with
$t_{min}=0\ (0\%)$.
The rhombicuboctahedron topology converges to the octahedron topology
with $t_{max}=u/2\ (100\%)$.
The truncated cube topology converges to the cuboctahedron topology
with $t_{max}=u/2\ (100\%)$.
Such parameters as the truncation
are not commonly controlled in other lattice \hl{modelling} tools such as
Autodesk Netfabb and nTopology.

Potentially, the proposed approach can reinforce other existing lattice
\hl{modelling} tools by introducing an additional F-rep tool to vary
geometric parameters\hl{,} which include and are not limited by the lattice
thickness.

\section{Implementation}
\label{sec:implementation}

For the proposed work, it was decided to use Open CASCADE Technology (OCCT)
which is the most widely used and well-documented open-source
GMK~\parencite{yuan2008development, Banovic2018}.
Developing the software prototype in the C++ programming language (which is
\hl{the native programming language of} OCCT) can \hl{undoubtedly} prove
to be useful in the \hl{long-run} for \hl{geometric modelling}
applications\hl{,} which is \hl{confirmed} by extensive use of C++ in every major existing CAD
software~\parencite{Li2011, golovanov2014geometric}.
However, for prototyping purposes, it was decided to develop a minimal viable
product (MVP) that would be faster to \hl{create} and easier to iterate for further
improvements~\parencite{Ries2011, Lenarduzzi2016}.
For this purpose, it was decided to build the software prototype based on CadQuery
\hl{as the modelling} tool and CQ-editor \hl{as the} graphical user interface
(GUI) shell, both licensed under a permissive license
and written in Python~\parencite{Urbanczyk2021, Urbanczyk2021b}.
CadQuery introduces \hl{parametrisation} methods built on top of OCCT\hl{, which}
are critical for the proposed work.
Moreover, OCCT and CadQuery are cross-platform, meaning
\hl{that the developed prototype has no limitation in terms of an OS it can
be used on}.

\hl{The development was focused only on the implementation of the methodology
itself.
This methodology is implemented in a Python library, which describes
all the geometry and constraints, and works together
with CadQuery.
The GUI is inherited from CQ-editor with an ability to import the
developed library from within the CQ-editor inline code editor.
The GUI also includes the 3D CAD viewer capable of rendering the resulting
solid models with the OCCT GMK.}

This section discusses the implementation of the proposed approach in detail.
Section~\ref{sec:imp-topologies} shows how function $T$ can be used to define
a lattice topology using the proposed \hl{method}.
Section~\ref{sec:imp-parameter-variation} provides examples of the
variation of geometric parameters $P$.
Section~\ref{sec:comp-performance} \hl{analyses} the computational performance
of the developed software prototype.

\subsection{Functional definition of lattice topologies}
\label{sec:imp-topologies}

The first step of the implementation is the definition of skeletal graphs of
topologies.
The implementation approaches differ between the beam-based and TPMS-based topologies.
However, in both cases, a function $T$ that defines the topology is required
to be \hl{determined}.

\subsubsection{Beam-based topologies}
\label{sec:imp-beams-framework}

For the beam-based topologies, the topologies are defined by the positions
of nodes and the line segments between them accordingly to each specific
topology.
\hl{The lines and nodes allow} to immediately obtain function $T$ as a
union of equations for multiple
straight lines.

\begin{figure*}[h!]
    \centering
    \begin{subfigure}{.23\linewidth}
        \centering
        \includegraphics[width=\textwidth]{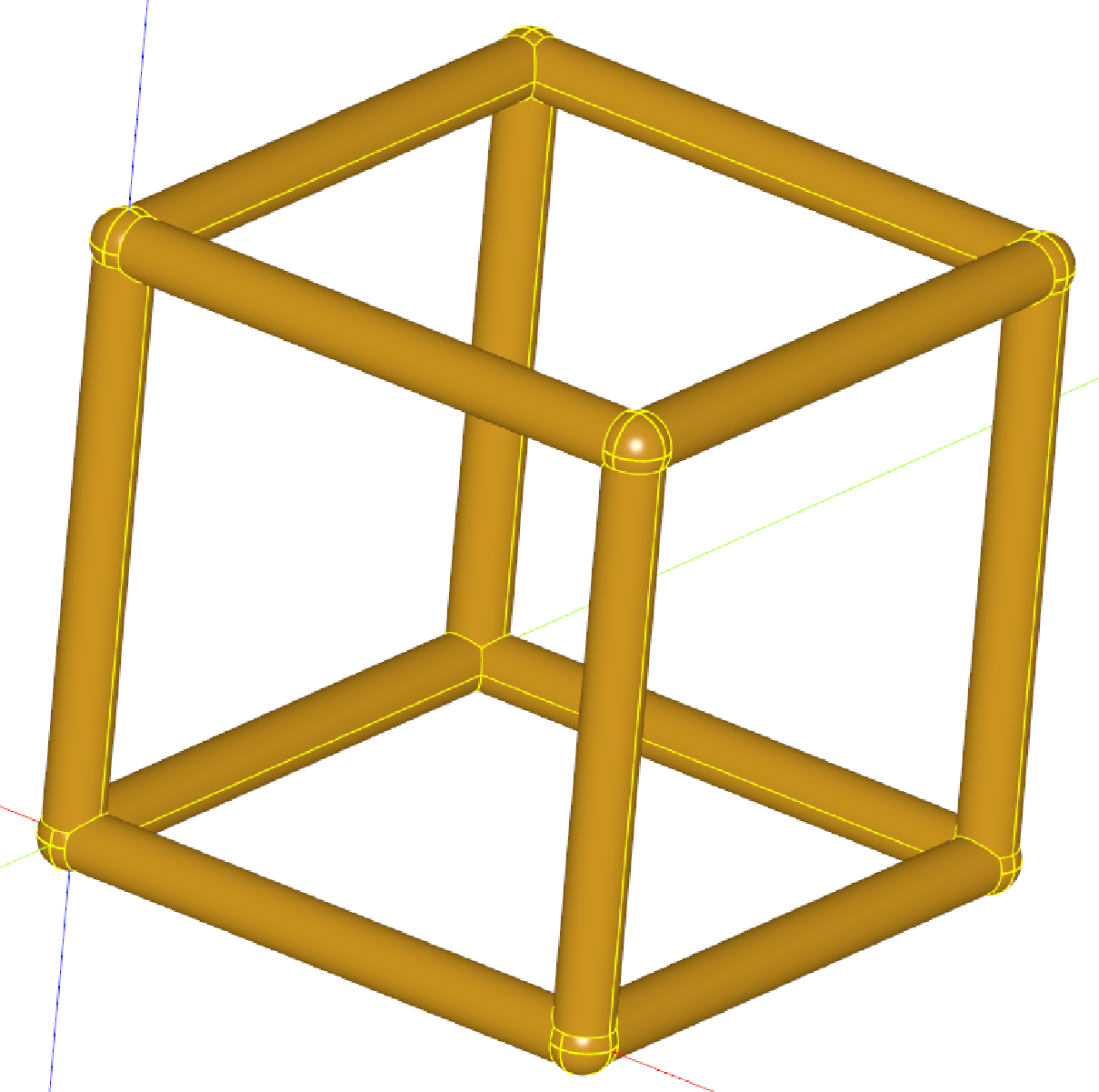}
        \caption{Simple cubic}\label{fig:cubic}
    \end{subfigure}
        \hfill
    \begin{subfigure}{.23\linewidth}
        \centering
        \includegraphics[width=\textwidth]{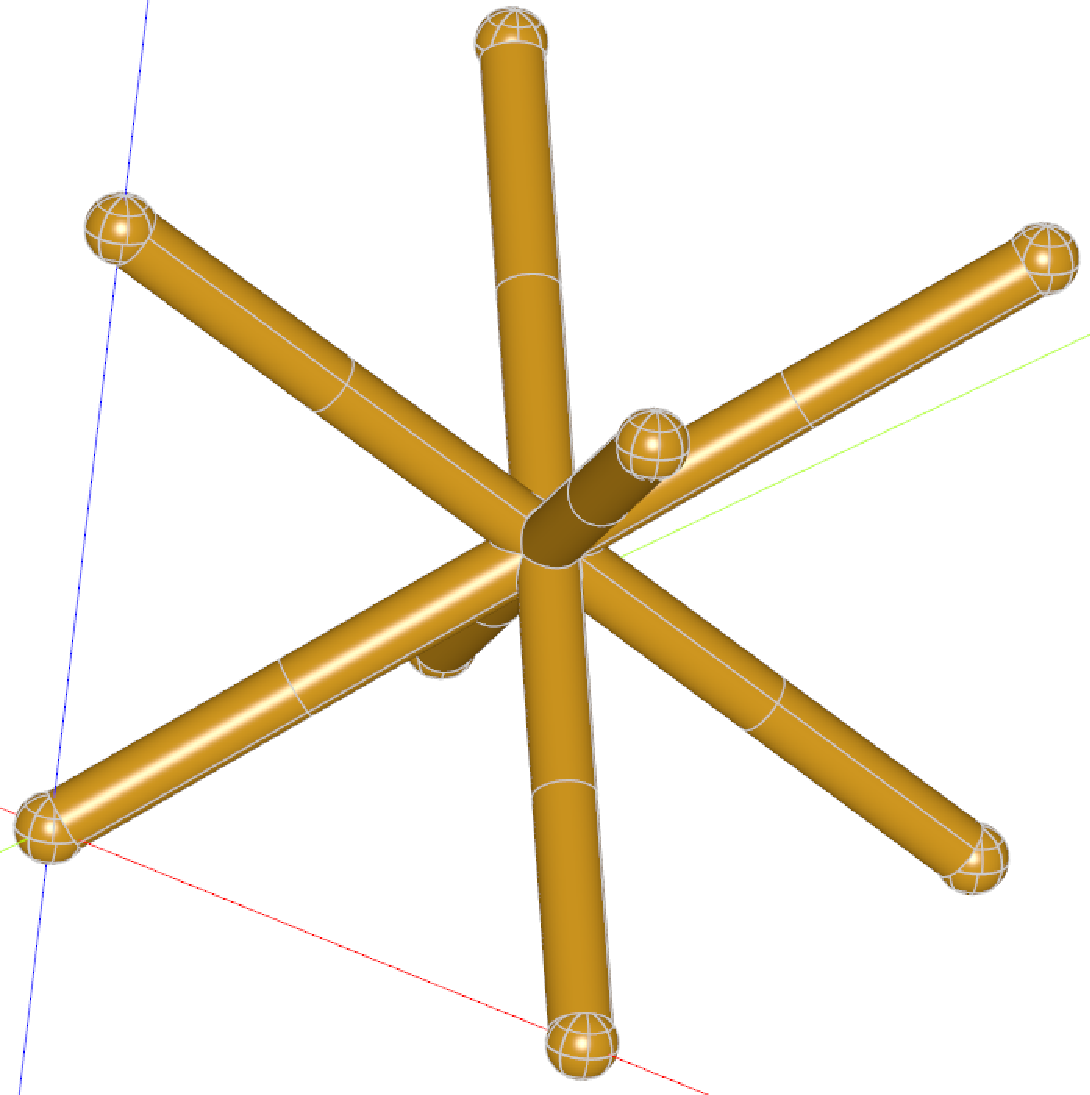}
        \caption{BCC}\label{fig:bcc}
    \end{subfigure}
       \hfill
    \begin{subfigure}{.23\linewidth}
        \centering
        \includegraphics[width=\textwidth]{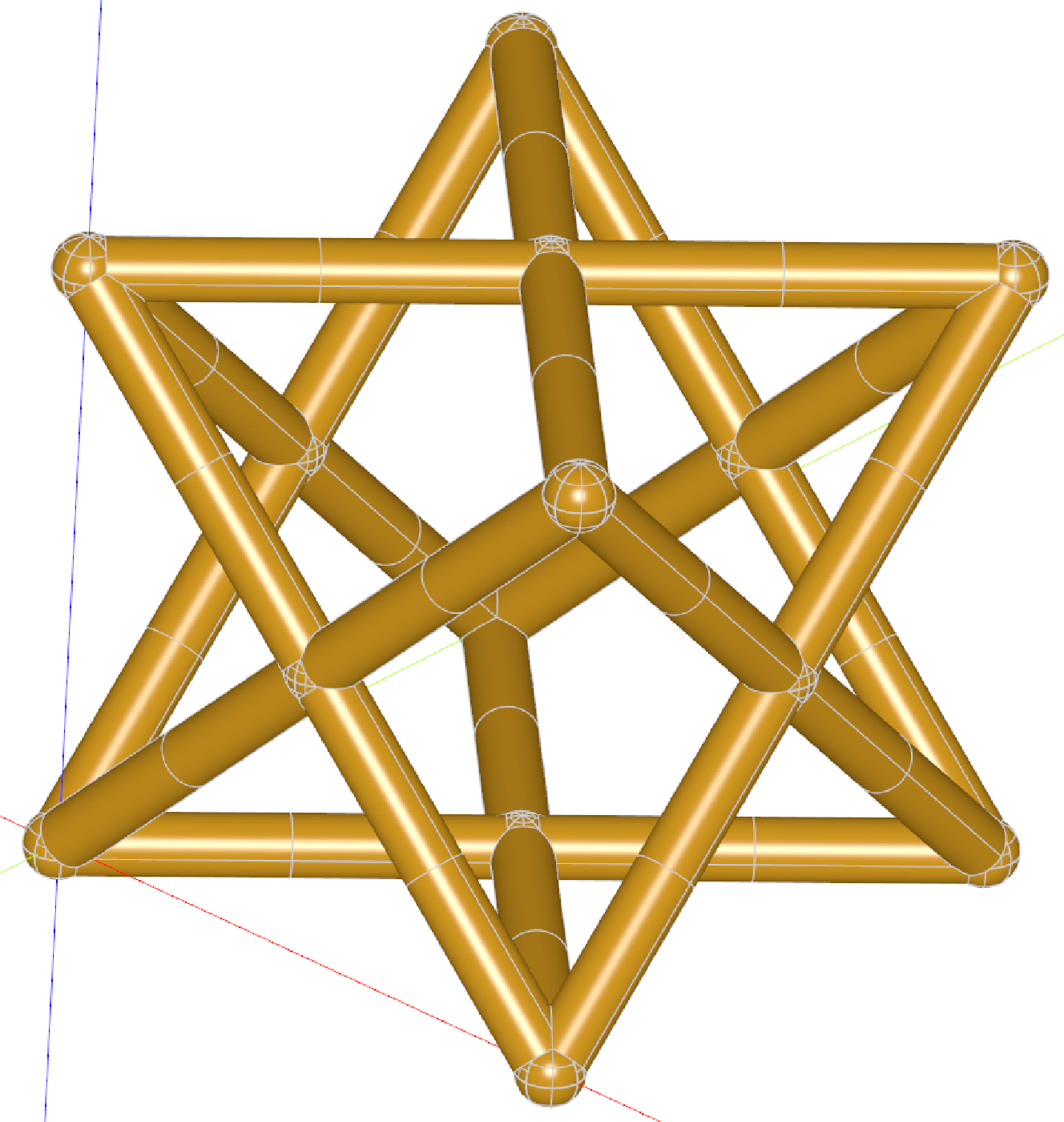}
        \caption{FCC}\label{fig:fcc}
    \end{subfigure}
        \hfill
    \begin{subfigure}{.23\linewidth}
        \centering
        \includegraphics[width=\textwidth]{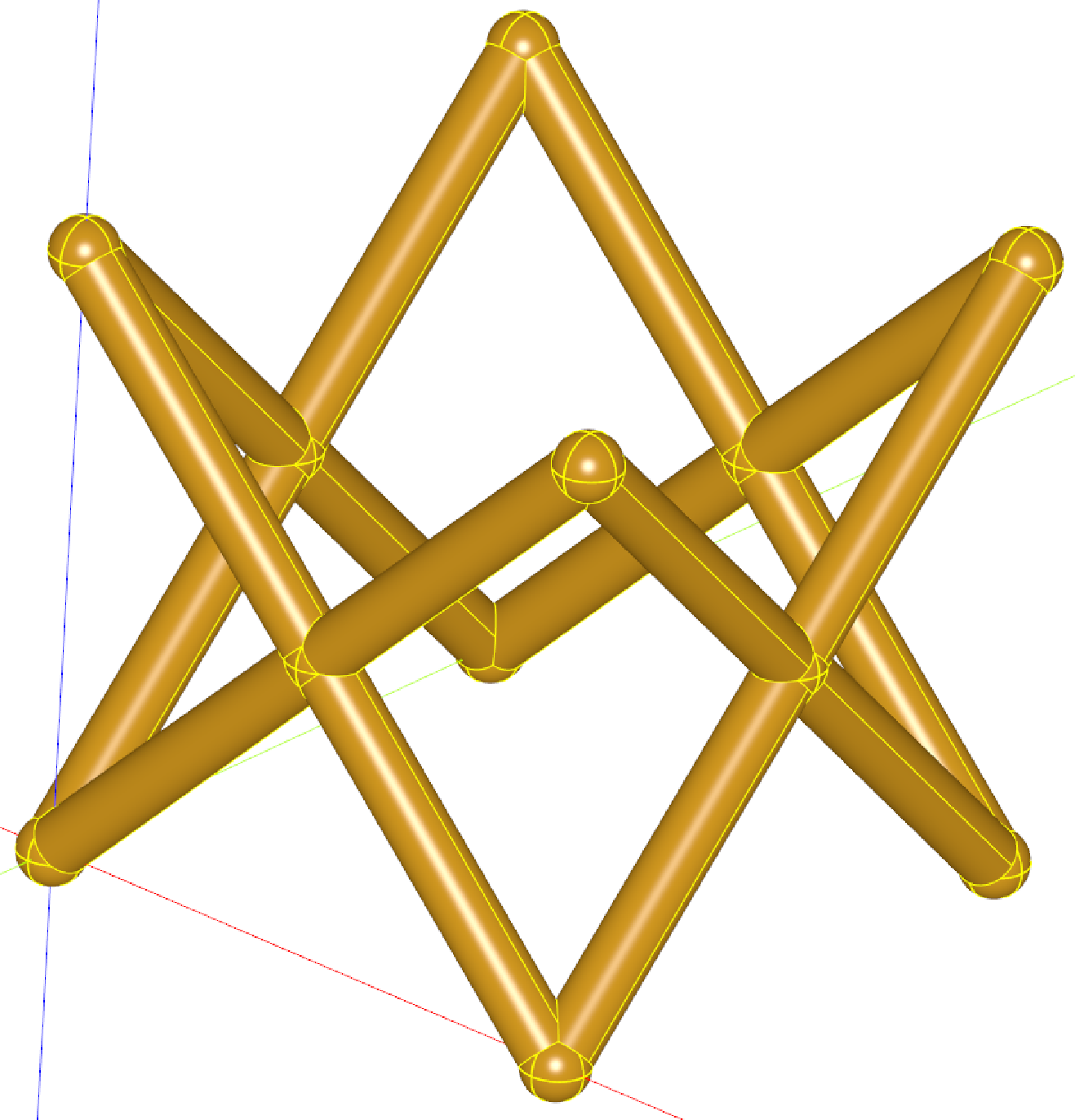}
        \caption{S-FCC}\label{fig:sfcc}
    \end{subfigure}
    
    \bigskip

    \centering
    \begin{subfigure}{.23\linewidth}
        \centering
        \includegraphics[width=\textwidth]{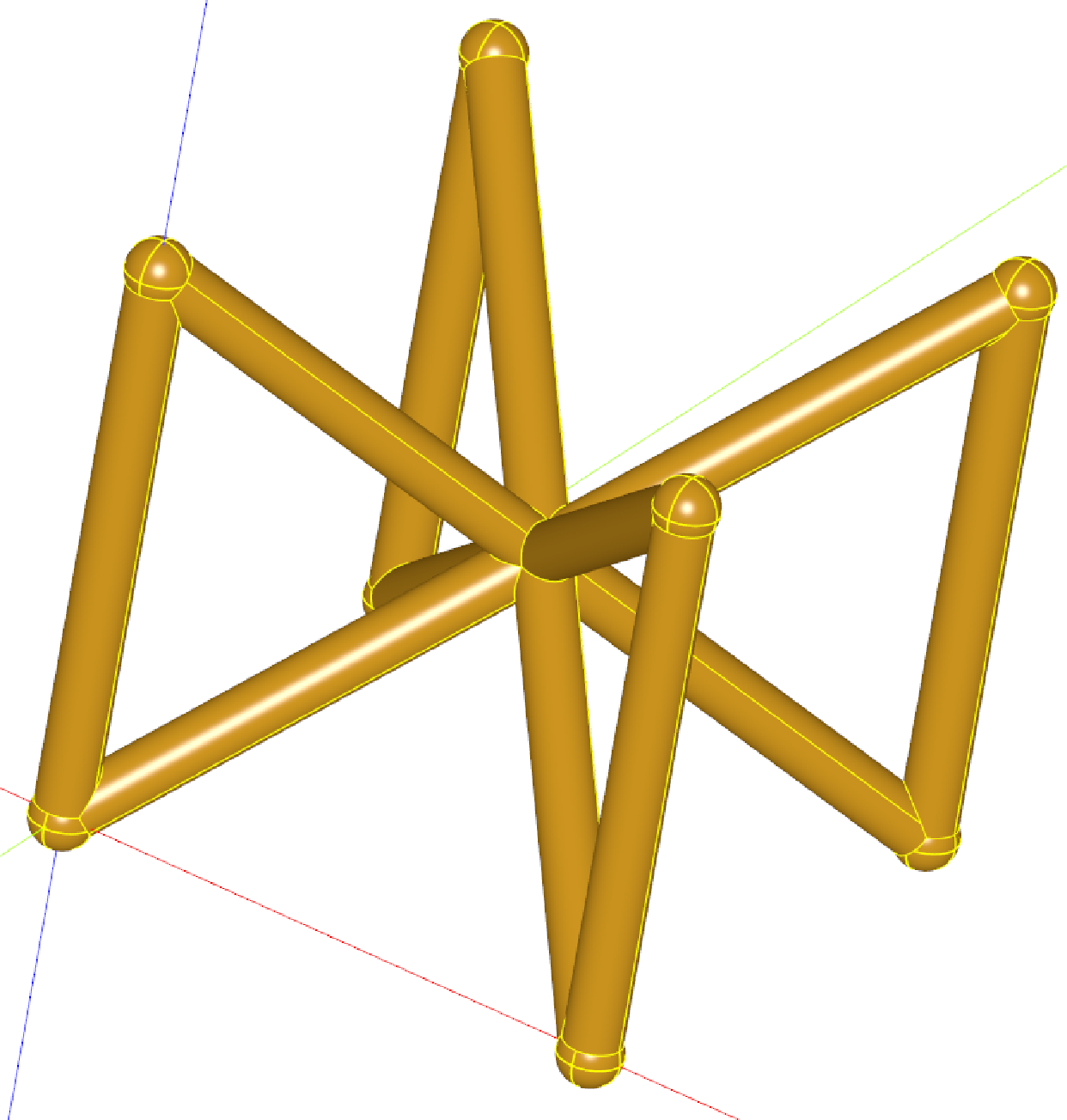}
        \caption{BCCz}\label{fig:bccz}
    \end{subfigure}
       \hfill
    \begin{subfigure}{.23\linewidth}
        \centering
        \includegraphics[width=\textwidth]{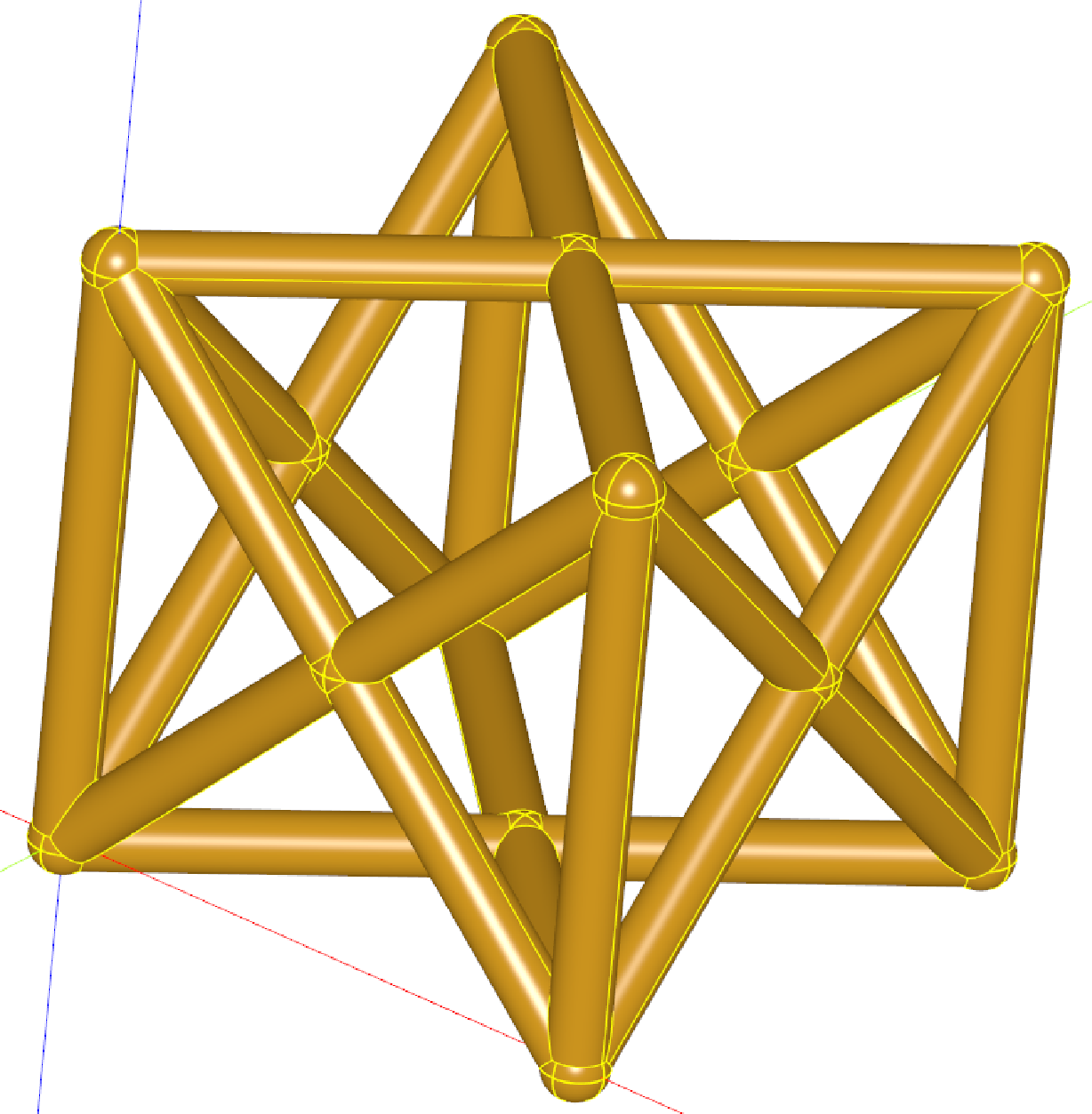}
        \caption{FCCz}\label{fig:fccz}
    \end{subfigure}
        \hfill
    \begin{subfigure}{.23\linewidth}
        \centering
        \includegraphics[width=\textwidth]{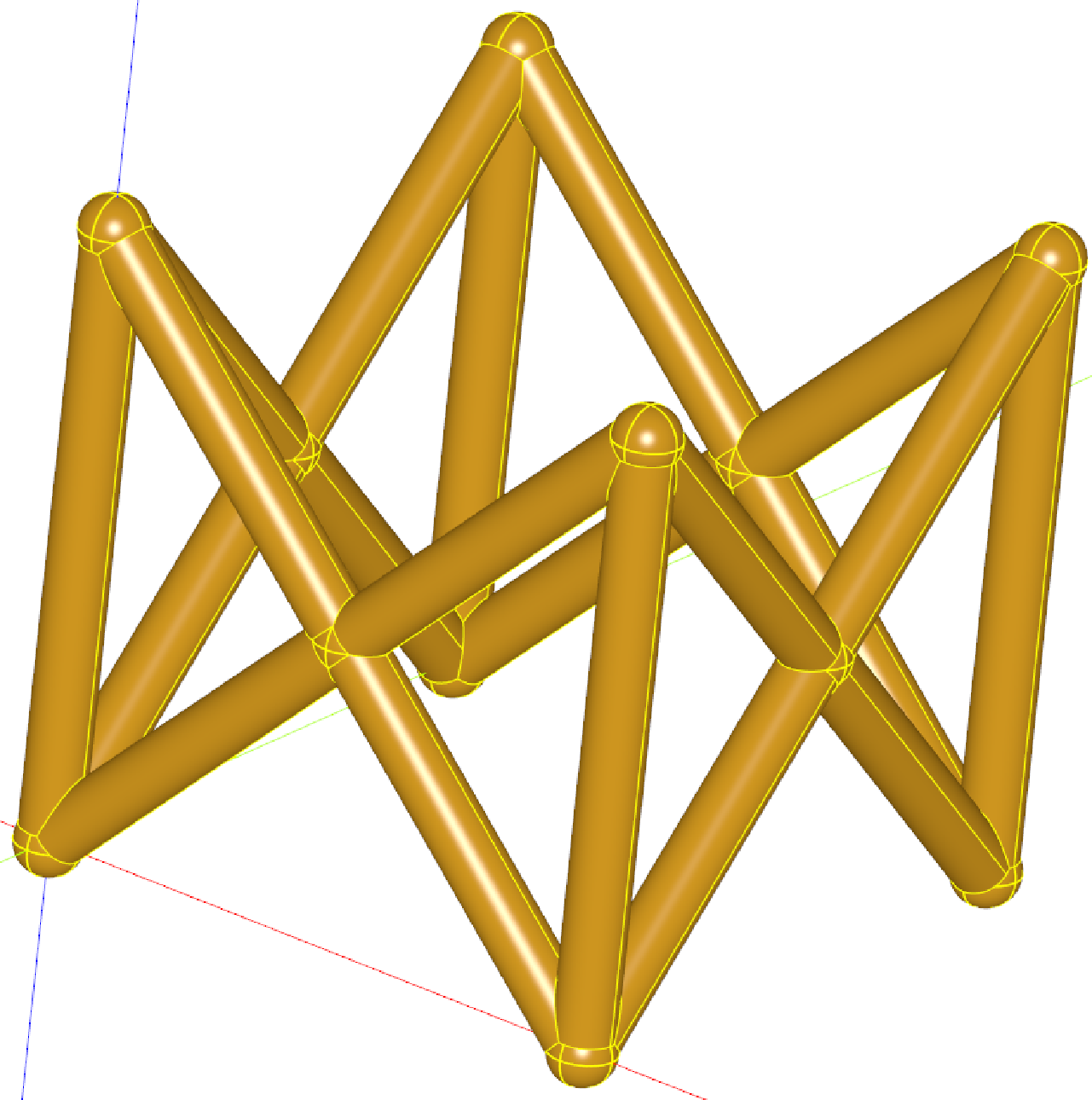}
        \caption{S-FCCz}\label{fig:sfccz}
    \end{subfigure}
        \hfill
    \begin{subfigure}{.23\linewidth}
        \centering
        \includegraphics[width=\textwidth]{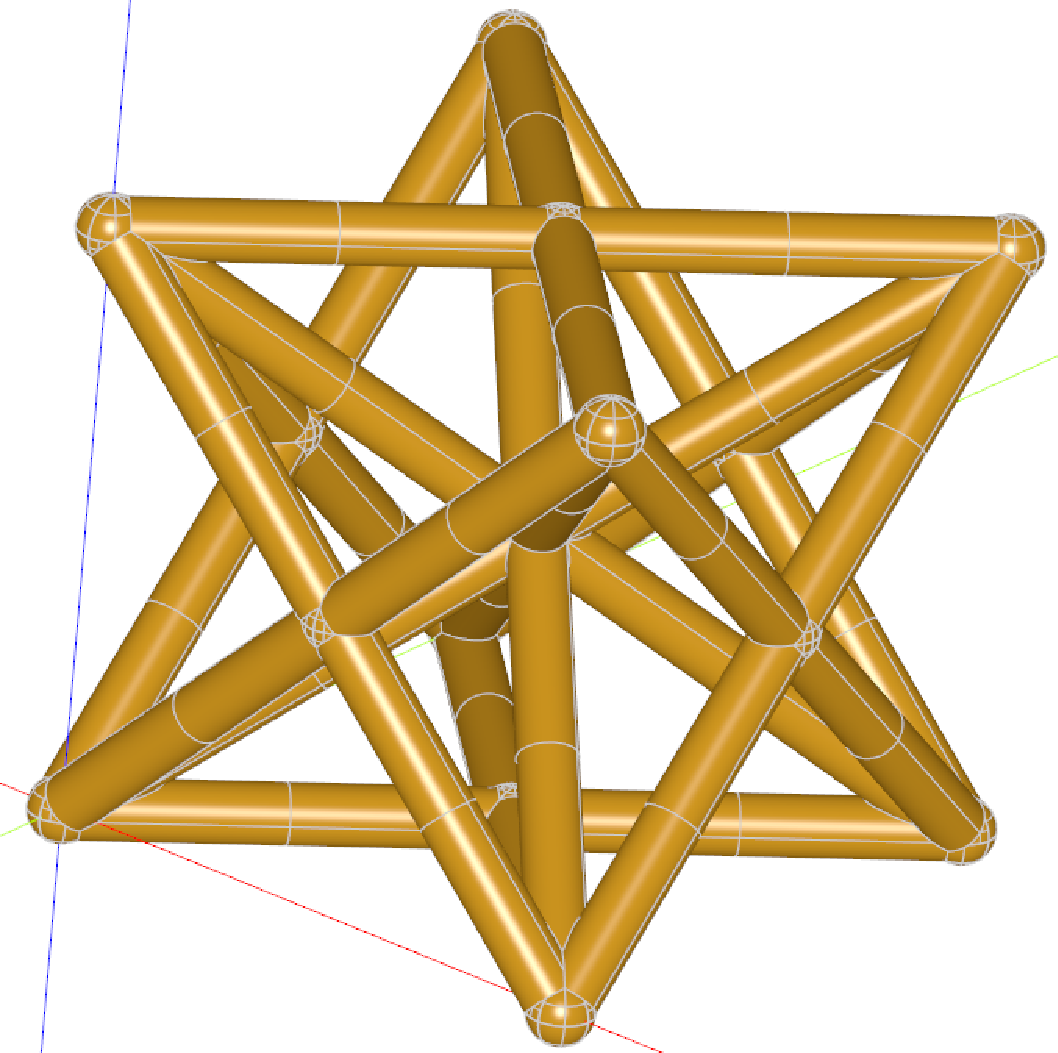}
        \caption{FBCC}\label{fig:fbcc}
    \end{subfigure}

    \bigskip

    \centering
    \begin{subfigure}{.23\linewidth}
        \centering
        \includegraphics[width=\textwidth]{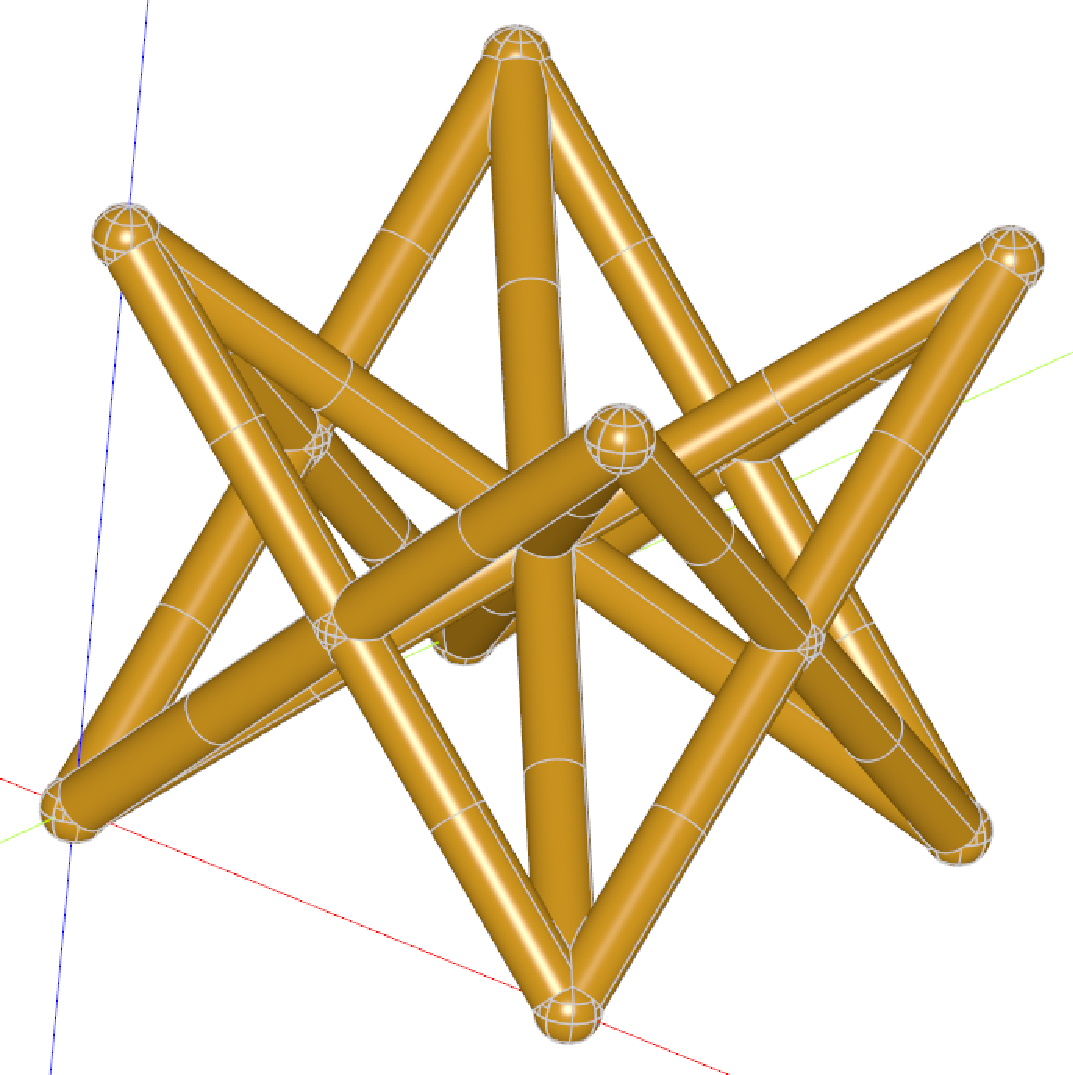}
        \caption{S-FBCC}\label{fig:sfbcc}
    \end{subfigure}
    \hspace{0.2\textwidth}
    \begin{subfigure}{.23\linewidth}
        \centering
        \includegraphics[width=\textwidth]{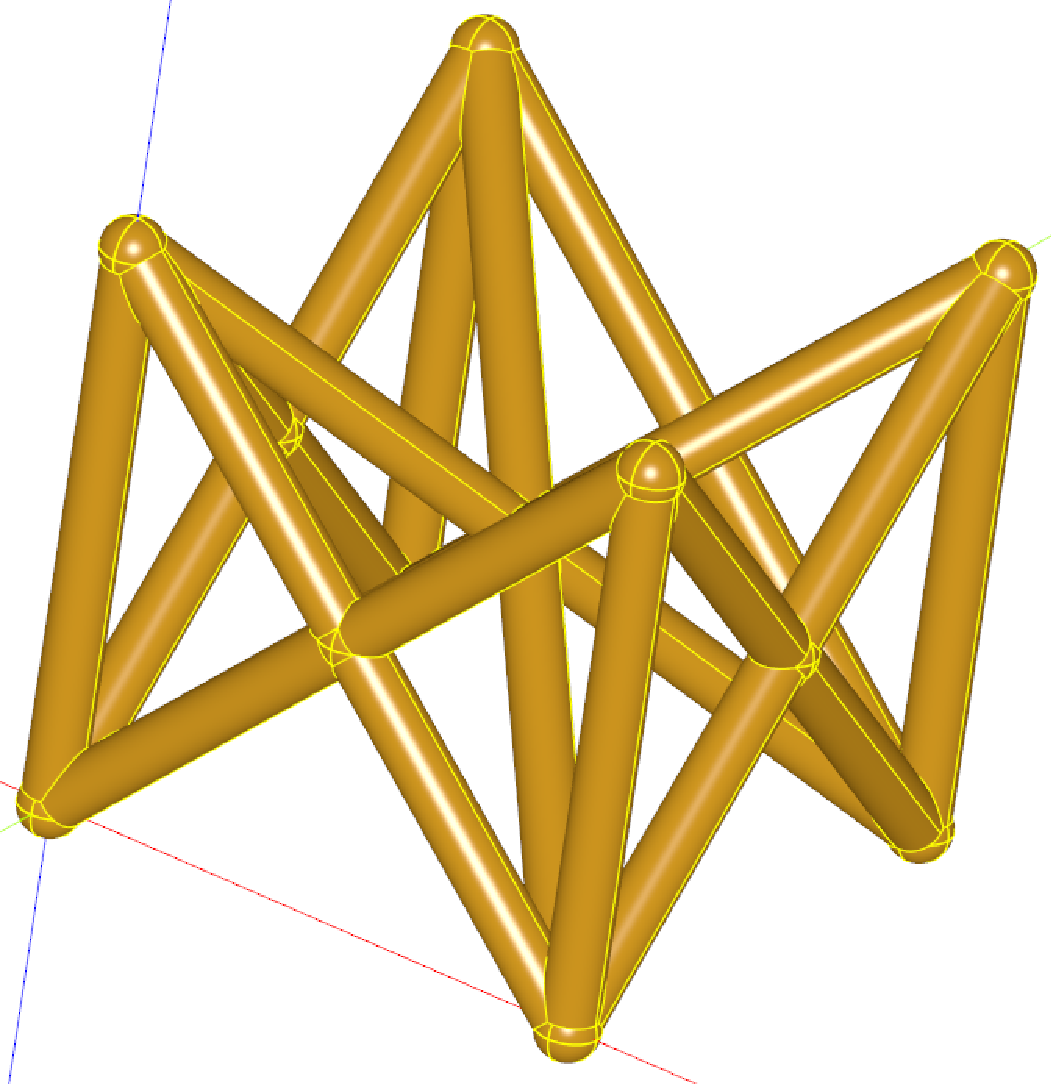}
        \caption{S-FBCCz}\label{fig:sfbccz}
    \end{subfigure}

    \caption{Various beam-based topologies
    inspired by the cubic crystal system
    supported by the
    developed approach}
    \label{fig:topologies}
\end{figure*}

\begin{figure*}[h!]
    \centering
    \begin{subfigure}{.23\linewidth}
        \centering
        \includegraphics[width=\textwidth]{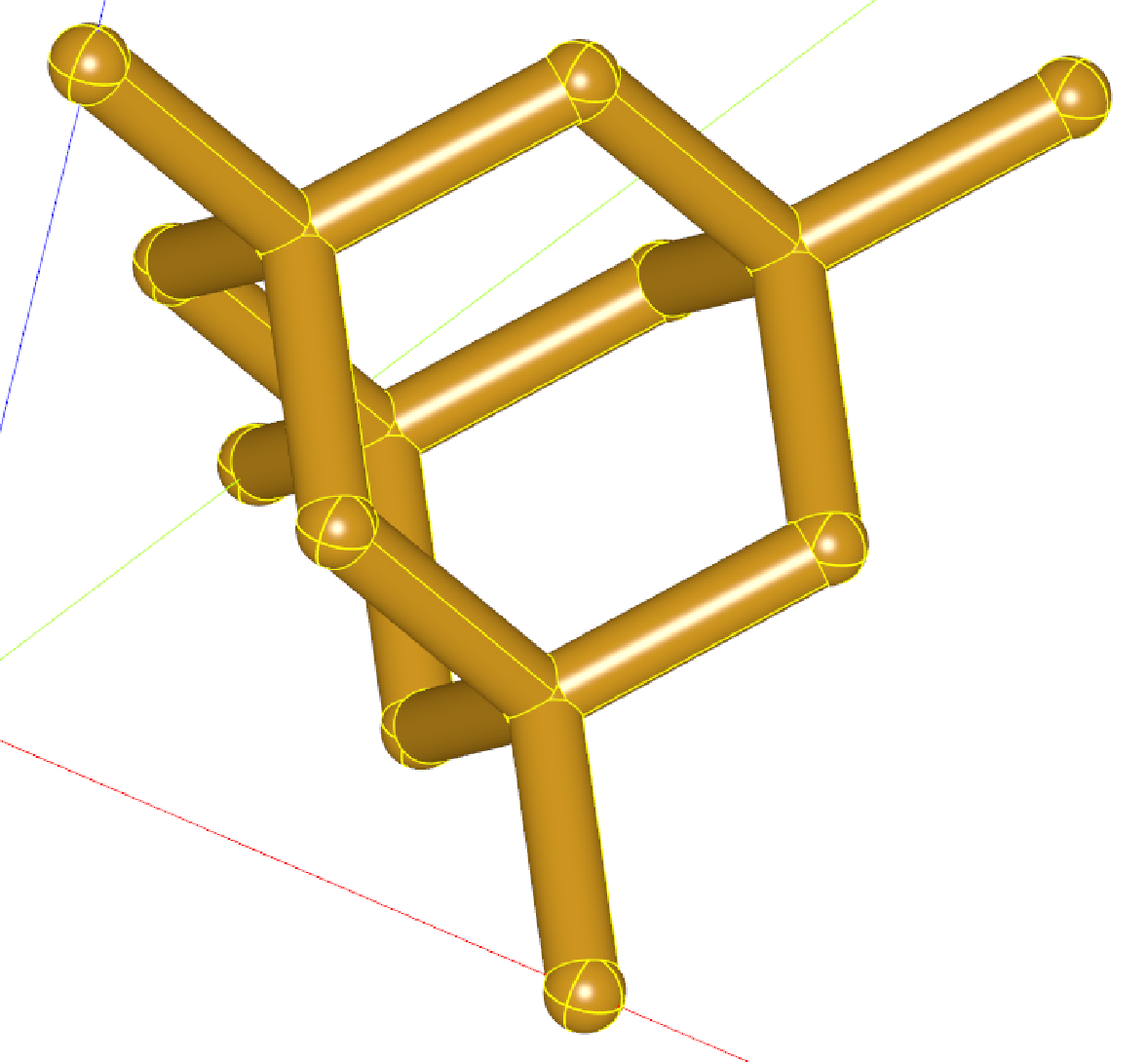}
        \caption{Diamond}\label{fig:diamond}
    \end{subfigure}
        \hfill
    \begin{subfigure}{.23\linewidth}
        \centering
        \includegraphics[width=\textwidth]{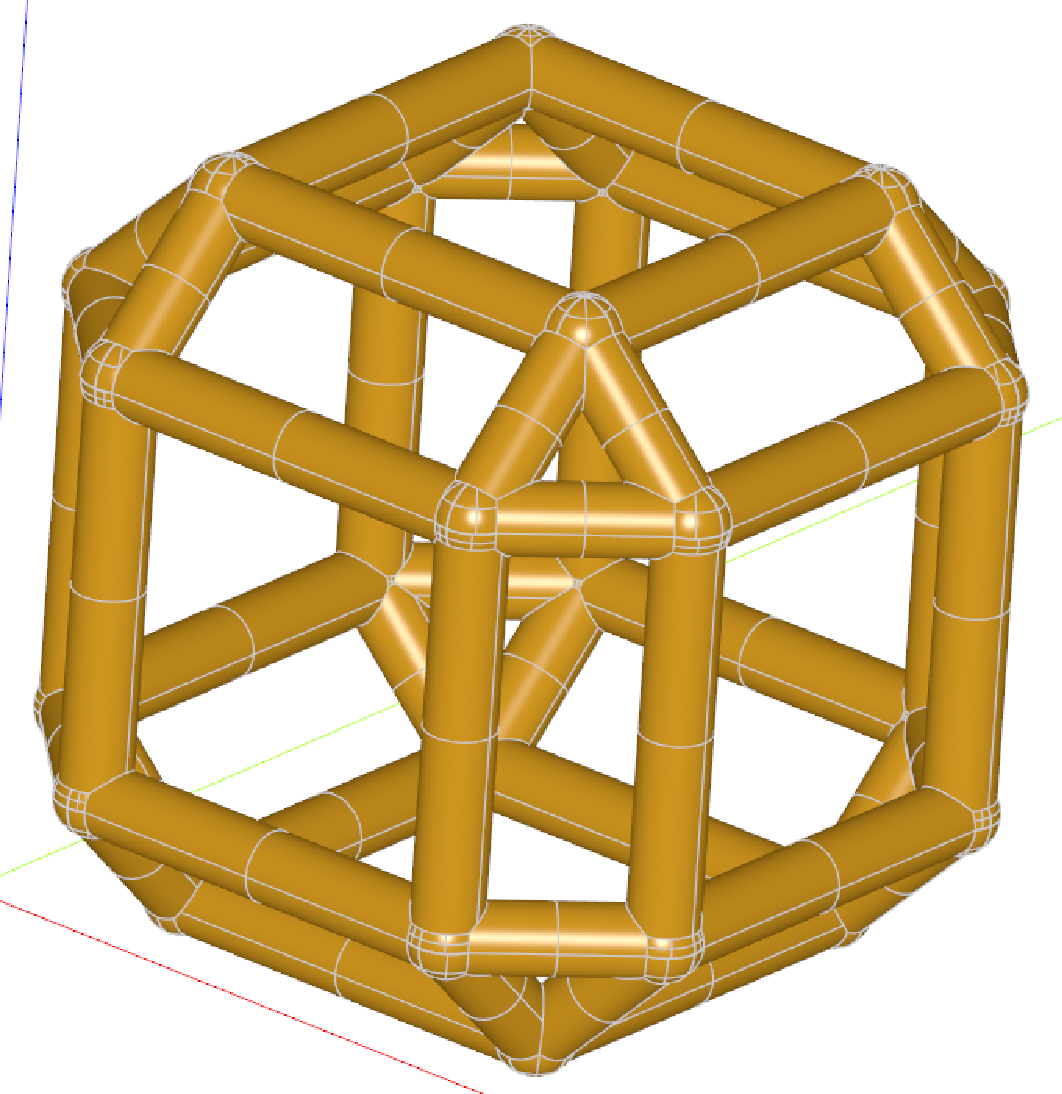}
        \caption{Rhombicuboctahedron}
        \label{fig:rco}
    \end{subfigure}
    \hfill
    \begin{subfigure}{.23\linewidth}
        \centering
        \includegraphics[width=\textwidth]{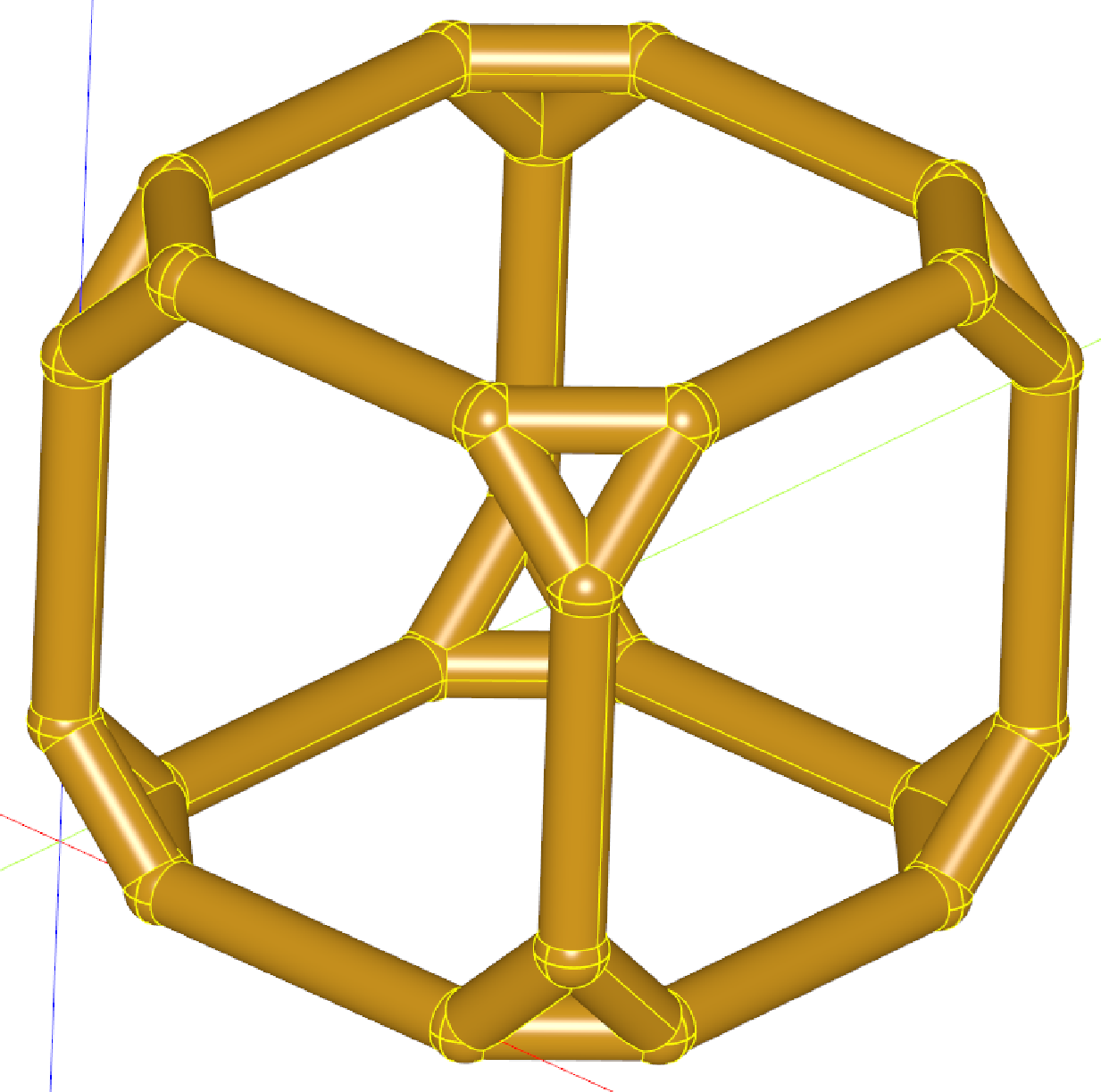}
        \caption{Truncated cube}
        \label{fig:t-cubic}
    \end{subfigure}

    \caption{Additional beam-based topologies supported by the
    developed approach}
    \label{fig:other-topologies}
\end{figure*}

Figure~\ref{fig:topologies} illustrates beam-based topologies
inspired by the cubic crystal system in crystallography that
the developed software prototype \hl{can} model, which includes:
simple cubic (Fig.~\ref{fig:cubic}),
\hl{body-centred} cubic (BCC) (Fig.~\ref{fig:bcc}), and
\hl{face-centred} cubic (FCC) (Fig.~\ref{fig:fcc}).
Several varieties of these topologies are supported as well, such as
self-supporting FCC without horizontal beams (S-FCC) (Fig.~\ref{fig:sfcc}),
BCC with additional 4 $z$-direction oriented beams (BCCz)
(Fig.~\ref{fig:bccz}),
FCC with additional 4 $z$-direction oriented beams (FCCz)
(Fig.~\ref{fig:bccz}),
S-FCCz (Fig.~\ref{fig:fccz}),
face- and \hl{body-centred} cubic (FBCC) (Fig.~\ref{fig:fbcc}),
S-FBCC (Fig.~\ref{fig:sfbcc})
S-FBCCz (Fig.~\ref{fig:sfbccz}).
All topologies shown in Fig.~\ref{fig:topologies}
have the cell size $u=10$ mm and
have 1 mm beam diameter and 1.1 mm node diameter.

Figure~\ref{fig:other-topologies} illustrates \hl{other}
supported beam-based topologies such as diamond
(Fig.~\ref{fig:diamond}),
rhombicuboctahedron
(Fig.~\ref{fig:rco}),
and truncated cube
(Fig.~\ref{fig:t-cubic}).
All topologies shown in
Fig.~\ref{fig:other-topologies}
have the cell size $u=10$mm and
have 1 mm beam diameter and 1.6 mm node diameter.
The truncation for the rhombicuboctahedron and the truncated
cube topologies in Fig.~\ref{fig:other-topologies} is 40\%.

\begin{figure*}[h!]
    \centering
    \begin{subfigure}{.23\linewidth}
        \centering
        \includegraphics[width=\textwidth]{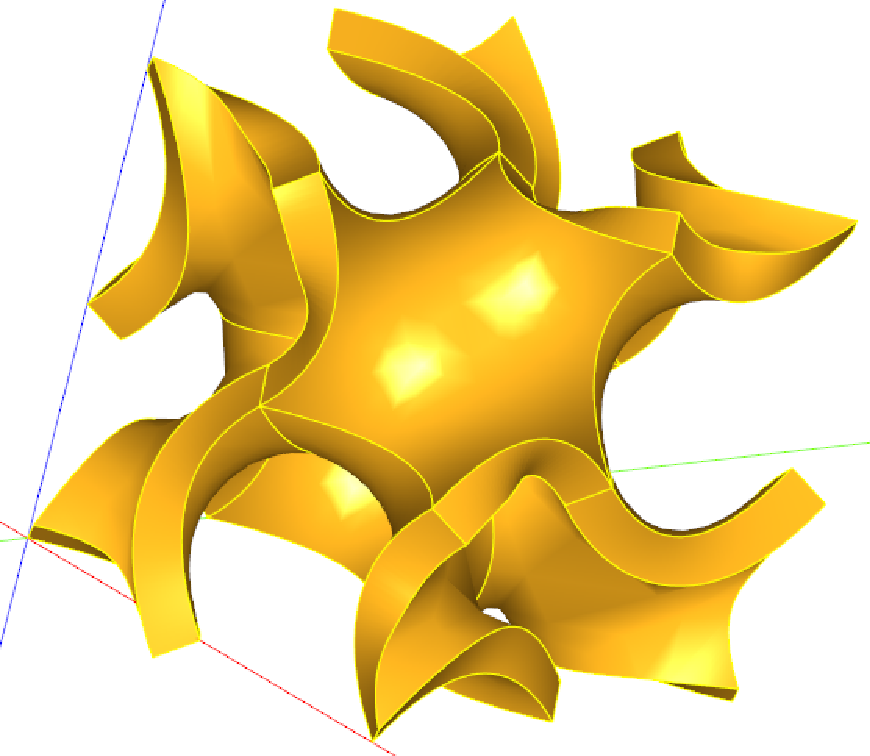}
        \caption{Gyroid}\label{fig:gyroid}
    \end{subfigure}
        \hfill
    \begin{subfigure}{.23\linewidth}
        \centering
        \includegraphics[width=\textwidth]{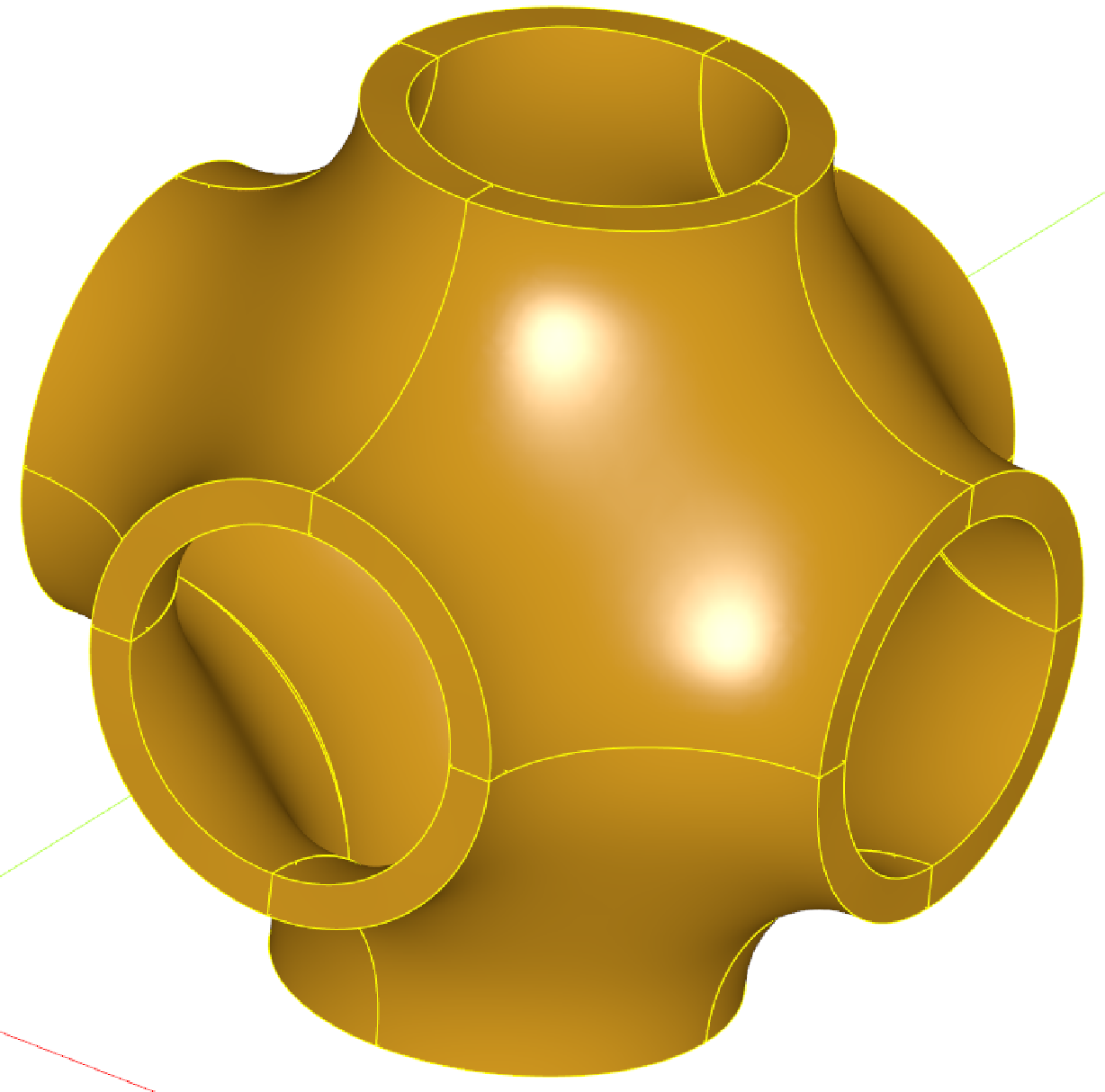}
        \caption{Schwarz P surface}\label{fig:schwartz}
    \end{subfigure}
        \hfill
    \begin{subfigure}{.23\linewidth}
        \centering
        \includegraphics[width=\textwidth]{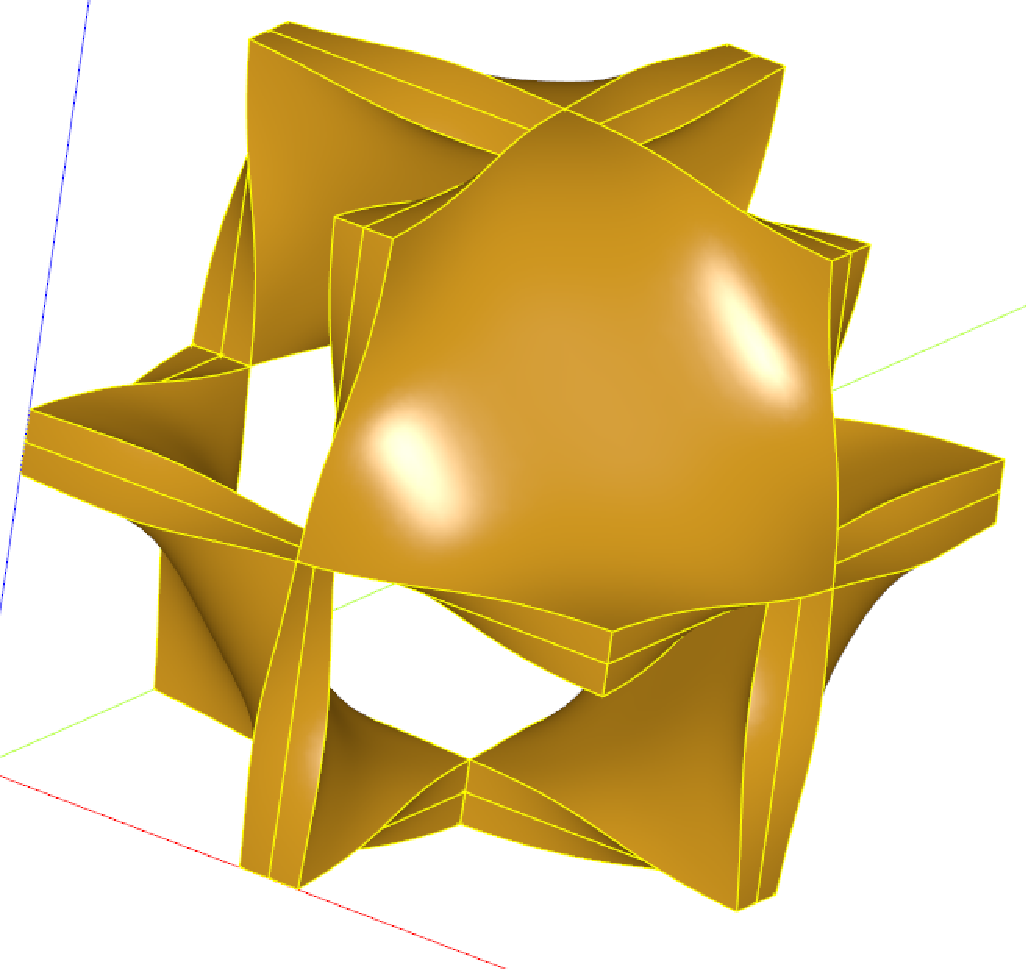}
        \caption{Schwarz D surface}\label{fig:schwartz-d}
    \end{subfigure}
    \caption{Various TPMS-based topologies supported by the
    developed approach}
    \label{fig:tpms-topologies}
\end{figure*}

The beam-based topologies
are common in design for AM (DFAM)~\parencite{Panesar2018},
but the process of their definition is based on F-rep in this work.
These topologies were defined in the software prototype while following
the object-oriented programming (OOP) principles which are crucial for any
CAD~\parencite{stroustrup1988object, warman1990object}.
In particular, certain features are repeated across topologies, such as,
for example, vertical $z$-oriented beams in simple cubic and BCCz topologies.
\hl{These features} were made into separate classes\hl{,} which are reused
in other topologies.
In the proposed work, OOP not only enables simple for the end-user modular
definition of topologies but also allows the end-user to define custom
topologies, i.e. the number of topologies possible to be \hl{modelled} is not
limited by the ones illustrated in Fig.~\ref{fig:topologies}.

\hl{The geometric modelling of beam-based topologies is implemented as follows.
A shape that defines the beam cross-section is defined.
For example, in Fig.}~\ref{fig:topologies}\hl{ and }Fig.~\ref{fig:other-topologies},
\hl{the shape of the cross-section is set to be a circle.
The skeletal graph in the case of the beam-based topologies is generated
as a wireframe
made of the instances of the }\verb|TopoDS_Wire|\hl{ class from OCCT.
The wires are then subject to the sweep operation, which is described by the}
\verb|BRepPrimAPI_MakeSweep|\hl{ class in OCCT, which generates a solid model
based on the wireframe, cross-section, and thickness.}

\hl{Special attention was dedicated to the ability to model the nodes of the
beam-based topologies.
Nodes are a critical component that allows a smooth transition between each unit cell
and enables better manufacturability of such lattice structures.}

\subsubsection{TPMS-based topologies}
\label{sec:imp-tpms-framework}

For the TPMS-based topologies,
array programming with the NumPy library is used~\parencite{harris2020array}.
In particular, NumPy allows the creation of linear spaces $x$, $y$, and $z$ and
\hl{uses} them as variables for any function.
Moreover, NumPy shows the extremely high performance when dealing with
large arrays of periodic data\hl{,} which are natural in \hl{the geometric
modelling of} lattice
structures since they are arrays themselves.
\hl{Thus, NumPy} opens a possibility to implement subdivision surfaces
for \hl{the modelling} of TPMS structures.
\hl{To achieve the effect of subdivision surfaces,} several sample points
\hl{are} taken from each octant of a unit cell
of a TPMS topology based on an implicit function that defines it.
It has been found that 18 points per octant (144 points per unit cell)
are sufficient \hl{to accurately represent}
the TPMS topologies supported by this approach.
These points are used for \hl{the modelling} of non-uniform rational B-splines
(NURBS) and surfaces of the skeletal graph based on them.
A \hl{modelling optimisation}
process was designed to model only an octant of a unit
cell and translate it \hl{seven times afterwards} to obtain \hl{an entire}
TPMS unit cell.
This \hl{optimisation} is seen as visible boundaries between each octant of
the unit cell in Fig.~\ref{fig:tpms-topologies}\hl{.}
The results of the mirroring are then united and
are considered a single solid model by Open CASCADE,
even though a boundary remains.
The definition of an octant rather than \hl{an entire}
unit cell is solely an \hl{optimisation} technique that works behind the scene
of the application.
\hl{Afterwards, the resulting surfaces need to be converted into solid objects.
The solidification is made possible by the implementation of the
}\verb|BRepOffsetAPI_MakeFilling|\hl{ class from OCCT, which allows the
generation of a solid object by offset from the NURBS surface.
This offset is one-directional in this class.
Thus, to generate a surface-based solid object with a thickness $t$,
two solid models are generated from a single surface: one with the $t/2$ offset
and another with the $-t/2$ offset.
This approach ensures that the surface is in the middle of the desired solid model.}

Figure~\ref{fig:gyroid}, Fig.~\ref{fig:schwartz},
and Fig.~\ref{fig:schwartz-d} illustrate the gyroid,
the Schwarz 'primitive' (P) surface, and the Schwarz 'diamond' (D) surface topologies,
respectively, \hl{modelled} with the proposed approach.
Table~\ref{tbl:tpms-topologies} covers 
the topology defining functions $T$ for these topologies.
The illustrated unit cells have the size $u=20$ and
the thickness $t=2$.
\hl{Note that there are visible boundaries between each octant of each
TPMS-based unit cell.
This effect appears due to the OOP optimisation mentioned above.
This optimisation technique models only one
octant and utilises it to build the rest of the unit cell.}

\subsection{Functional variation of geometric parameters}
\label{sec:imp-parameter-variation}

The next step of the implementation is to transform the obtained skeletal graph
into a solid body by adding thickness.
\hl{The beam-based topologies are solidified} by defining a cross-section
of the beam and its further application to each line segment \hl{that forms
the beam}.
\hl{The TPMS-based topologies are solidified by utilising}
the ability of OCCT to model solid bodies by B-rep offset.
Considering that the thickness of a TPMS-based unit cell of a lattice
structure is set to be $t$, a $t/2$ offset in both normal directions
needs to be applied to the NURBS
surface that forms the skeletal graph of the topology.

As for the implementation of the variation of geometric parameters
of a lattice structure, linear spaces generated by NumPy are used
as an input to any arbitrary function that defines the distribution of the
parameters.

As an example of the implementation,
consider a heterogeneous lattice based on the Schwarz P surface
defined according to the proposed method.
Let the unit cell size be set to 20 mm, the lattice size
to $10\times10\times10$, and the thickness
\hl{be linearly changing} from 0.1 mm to 7.0 mm
along the $z$-axis.
The Schwarz P surface itself is defined by its skeletal graph
\begin{equation}
    T(\mathbf{X}): \cos(x)+\cos(y)+\cos(z)=0,
    \label{eq:schwartz}
\end{equation}
i.e. by a TPMS with the zero thickness $t$~\parencite{Michielsen2008}.

\begin{figure*}
    \centering
    \begin{minipage}[c][16cm][t]{.7\textwidth}
        \centering
        \includegraphics[width=0.9\textwidth]{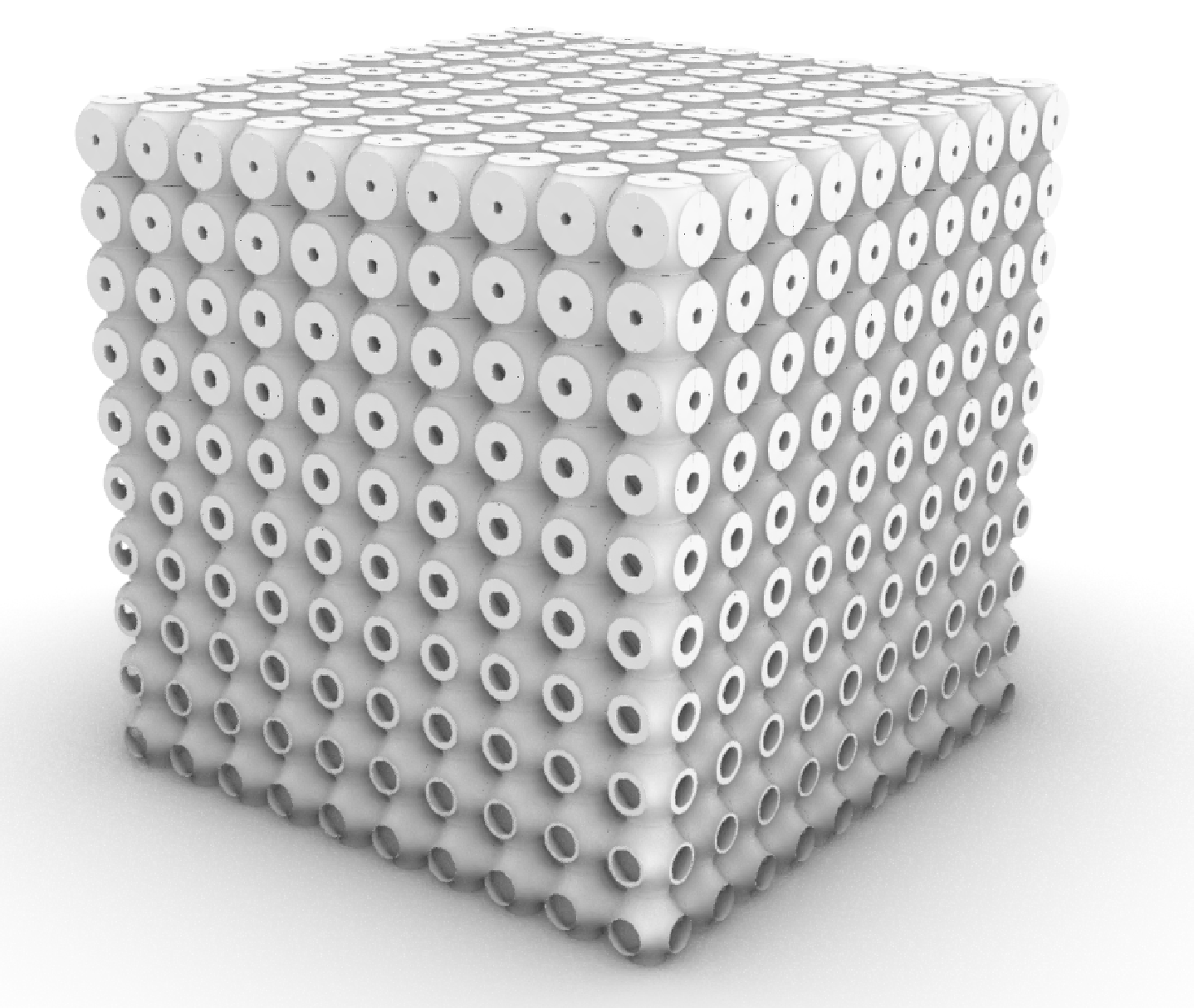}
      \subcaption{An isometric view on a heterogeneous
      lattice with the Schwarz P topology}
      \label{fig:hetero-schwartz-iso}
      \vspace*{\fill}
      \centering
      \includegraphics[width=0.9\textwidth]{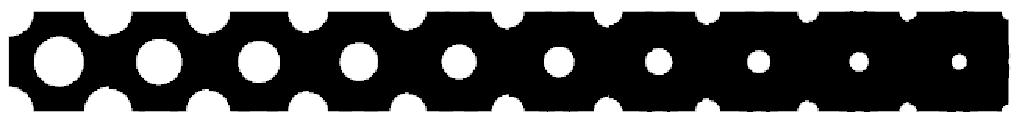}
      \subcaption{A side view on a single column in the $z$-direction
      of heterogeneous
      lattice with the Schwarz P topology}
      \label{fig:hetero-schwartz-side}\par\vfill
      \centering
      \includegraphics[width=0.9\textwidth]{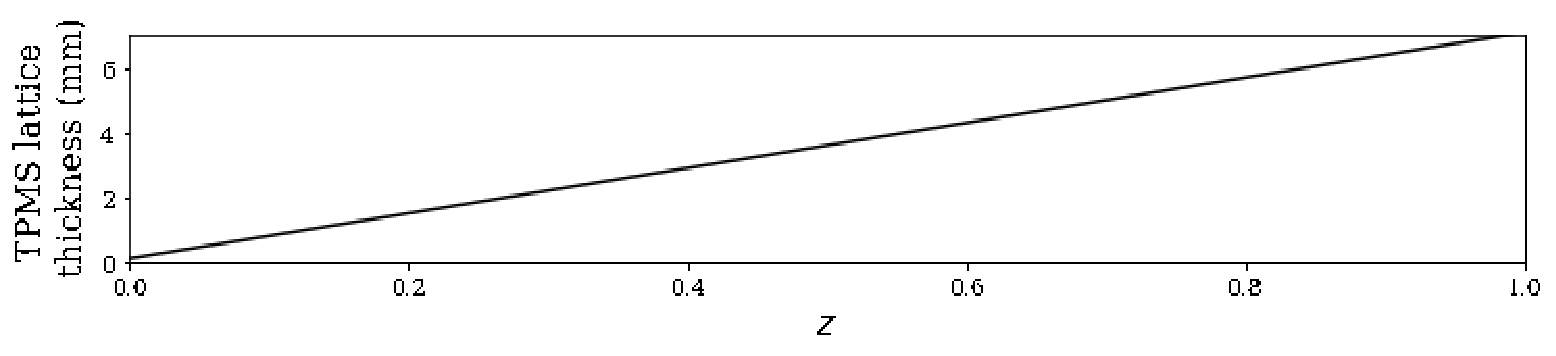}
      \subcaption{The linear function
      $P$ that corresponds to the thickness of the TPMS-based
      lattice vs the $z$ coordinate}
      \label{fig:line}
    \end{minipage}
    \caption{A heterogeneous lattice structure with the Schwarz P surface
    topology with linearly varying thickness generated with the proposed approach}
    \label{fig:hetero-schwartz}
\end{figure*}

F-rep can be overly complicated when used by an engineering designer as
a \hl{modelling} tool, as it requires expertise in both design and mathematics.
To simplify and adjust F-rep to the AM needs, the solid model\hl{,} which is based
on a skeletal graph\hl{,} can be obtained
by an additional function developed specifically
for this topology.
For example, for the Schwarz P surface topology, this additional function appears
as a single function with multiple parameters in it:
\begin{equation}
    \text{Schwarz lattice}\left(\text{size}\left(\mathbb{U}\right), N_x, N_y, N_z, P\left(\mathbf{X}\right)\right),
\end{equation}
where $\text{size}(\mathbb{U}) = 10$ mm is the size of the unit cell,
$N_x=10$, $N_y=10$, $N_z=10$ are the numbers of unit cells in
$x$, $y$, and $z$ directions, respectively,
and $P(\mathbf{X})$ is the function that controls the 
thickness of the TPMS-based structure, thus enabling setting $t>0$ and  construction
of solids based on surfaces.
$P(\mathbf{X})$ in this case corresponds to
\begin{equation}
    P(\mathbf{X}): t(z) = 6.9z + 0.1,
\label{eq:linear}
\end{equation}
where $t$ is the thickness of the lattice\hl{,} and $z\in[0, 1]$ is the variable
corresponding to the $z$-axis.
Note that since $z\in[0, 1]$, it has to be mapped to the actual
coordinate $z_a\in[1, N_z]$
with $z_a\in\mathbb{N^+}$.
The resulting solid model is a heterogeneous Schwarz P lattice with \hl{the}
varying thickness and is illustrated in Fig.~\ref{fig:hetero-schwartz}.

\begin{figure*}[h!]
    \centering
    \begin{minipage}[c][13cm][t]{.8\textwidth}
      \centering
      \includegraphics[width=0.9\textwidth]{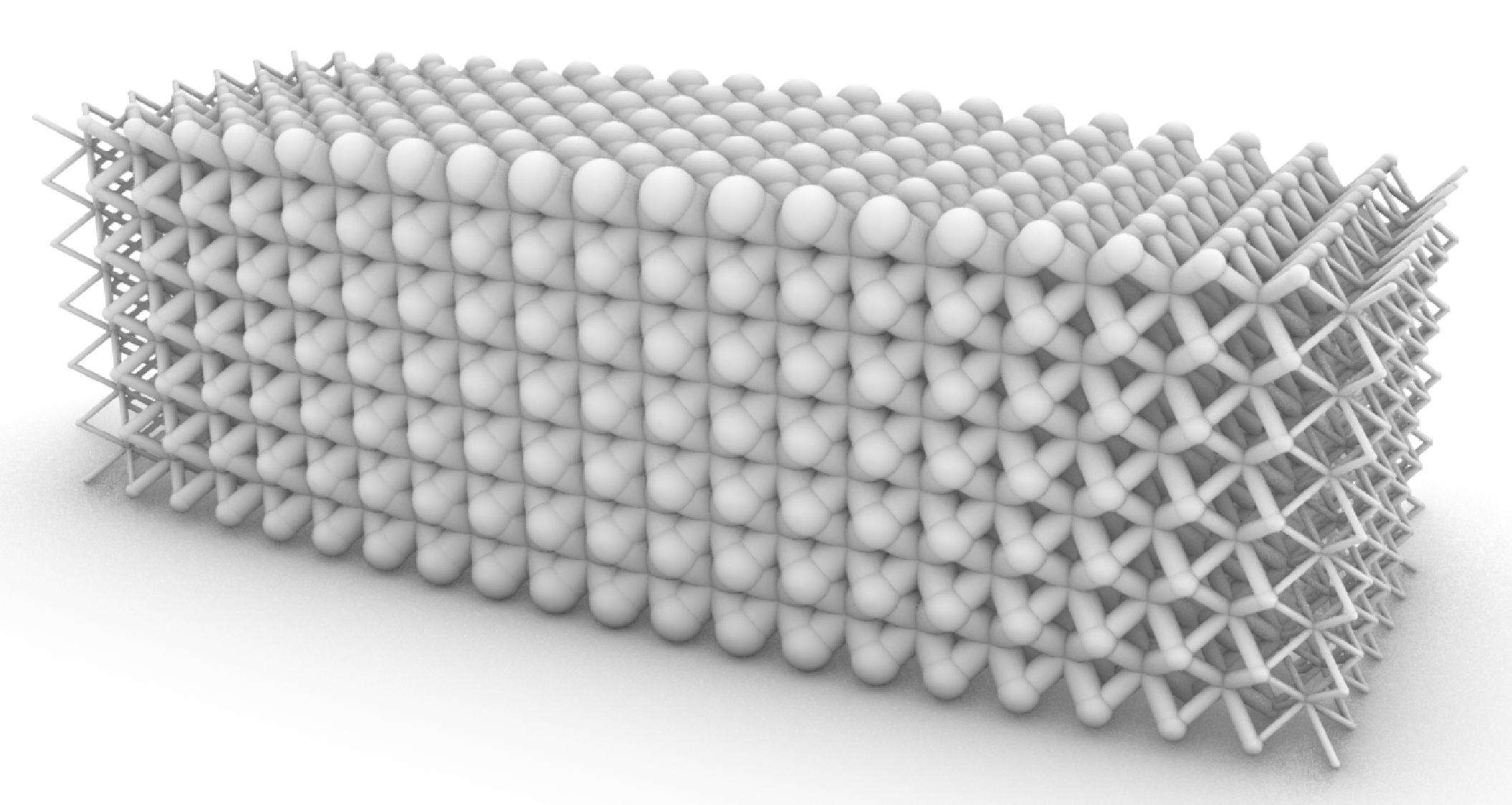}
      \subcaption{An isometric view on a heterogeneous
      lattice with the BCC topology}
      \label{fig:bcc-parabolic-iso}
      \centering
      \includegraphics[width=0.8\textwidth]{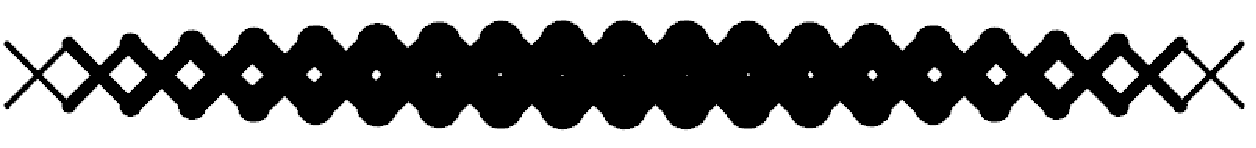}
      \subcaption{A side view on a single column in the $x$-direction
      of heterogeneous
      lattice with the BCC topology}
      \label{fig:bcc-parabolic}\par\vfill
      \centering
      \includegraphics[width=0.9\textwidth]{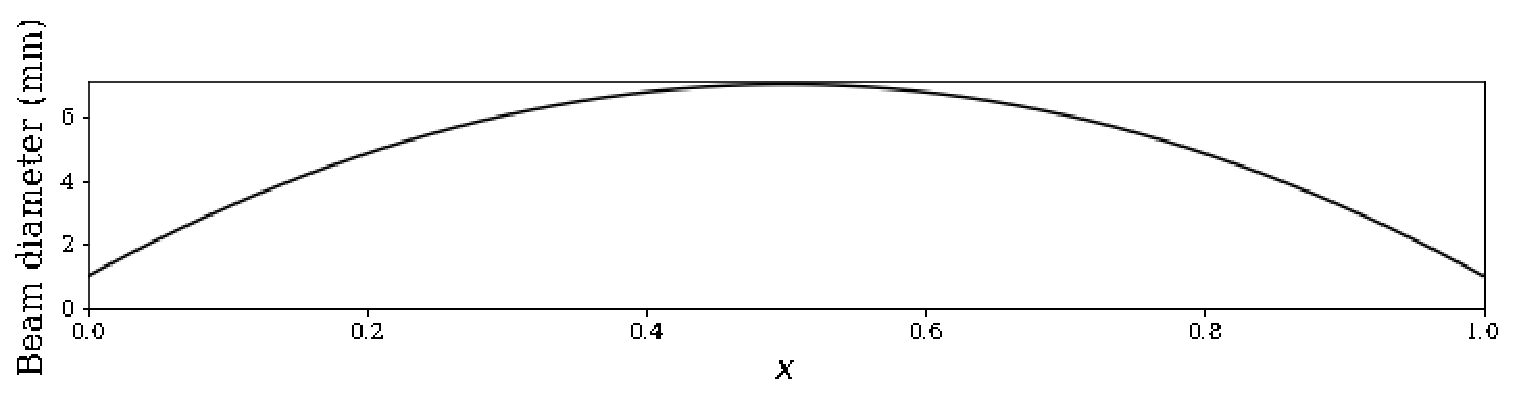}
      \subcaption{The parabolic function
      $P$ that corresponds to the thickness of the beam-based
      lattice vs the $x$ coordinate}
      \label{fig:parabola}
    \end{minipage}
    \caption{A heterogeneous lattice structure with the BCC
    topology with non-linearly varying thickness generated with the proposed approach}
    \label{fig:bcc-fun}
\end{figure*}

$P(\mathbf{X})$ is not limited to \hl{being} linear as in Equation~\ref{eq:linear}.
For example, consider a BCC lattice structure
with the size of $N_x=20$, $N_y=6$, $N_z=6$ with the beam diameter
controlled by the parabolic function:
\begin{equation}
    P(\mathbf{X}): D(x) = -4D_{\text{max}}(x-0.5)^2+D_{\text{max}}+D_{\text{min}},
    \label{eq:parabola}
\end{equation}
where $D$ is the diameter of a beam of the BCC lattice,
and $D_{\text{min}}$ and $D_{\text{max}}$ are the minimal and the
maximal diameters of the lattice, respectively.
This function was chosen as an example since it is symmetrical
around $x=0.5$.
For this example, $D_{\text{min}}=1$ mm and $D_{\text{max}}=6$ mm
were selected.
The resulting BCC structure is illustrated in Fig.~\ref{fig:bcc-fun}.
\hl{The lattice} has its beam diameter varying along the \hl{$z$-axis,}
which follows the parabolic function in Equation~\ref{eq:parabola}.
Observe that the node diameter varies since it is also a parameter that
can be controlled by a function and the process is virtually the same.
In this particular example, the node diameter was chosen to be 10\% larger
than the beam diameter.

Non-linear variation of geometric parameters of lattice structures is
not present \hl{in} other existing tools such as Autodesk
Netfabb\hl{,} which are limited to a linear change of parameters.

\begin{figure*}[h!]
    \centering
    \begin{minipage}[c][19cm][t]{.7\textwidth}
      \centering
      \includegraphics[width=0.7\textwidth]{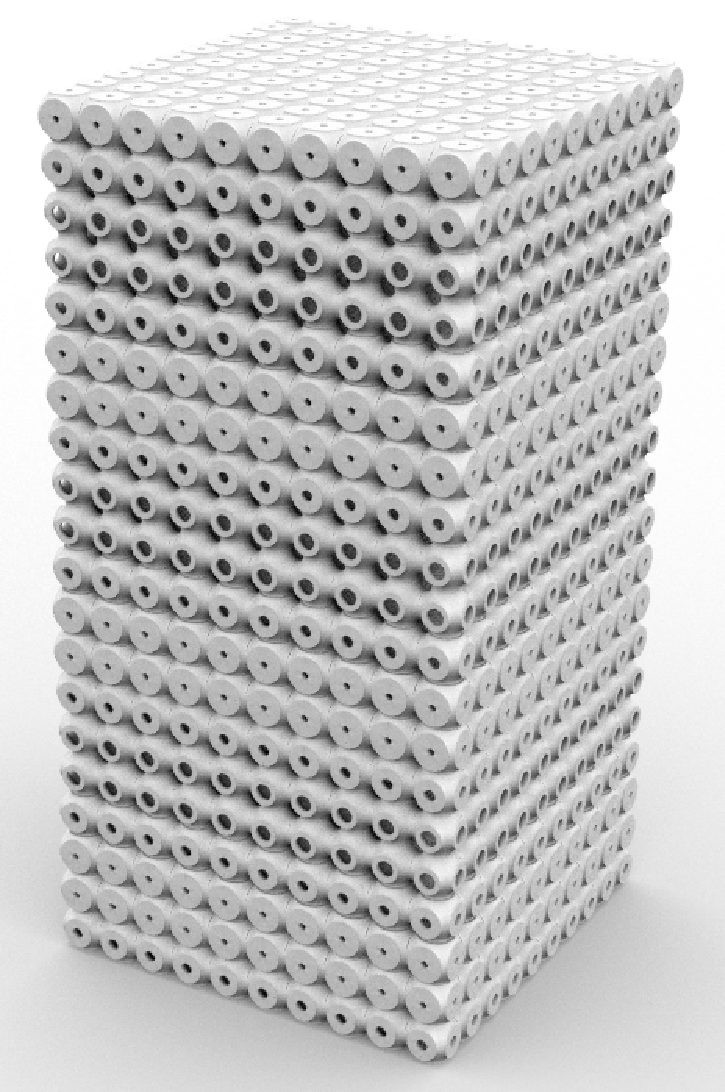}
      \subcaption{An isometric view on a heterogeneous
      lattice based on the Schwarz P surface}
      \label{fig:schwartz-sin-iso}
      \centering
      \includegraphics[width=0.7\textwidth]{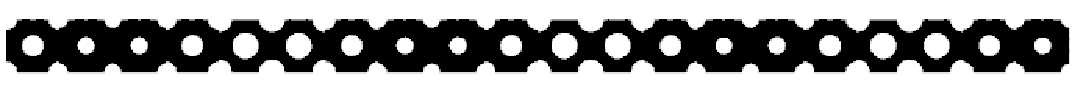}
      \subcaption{A side view on a single column in the $z$-direction
      of heterogeneous
      lattice with the Schwarz P surface topology}
      \label{fig:schwartz-sin}\par\vfill
      \centering
      \includegraphics[width=0.9\textwidth]{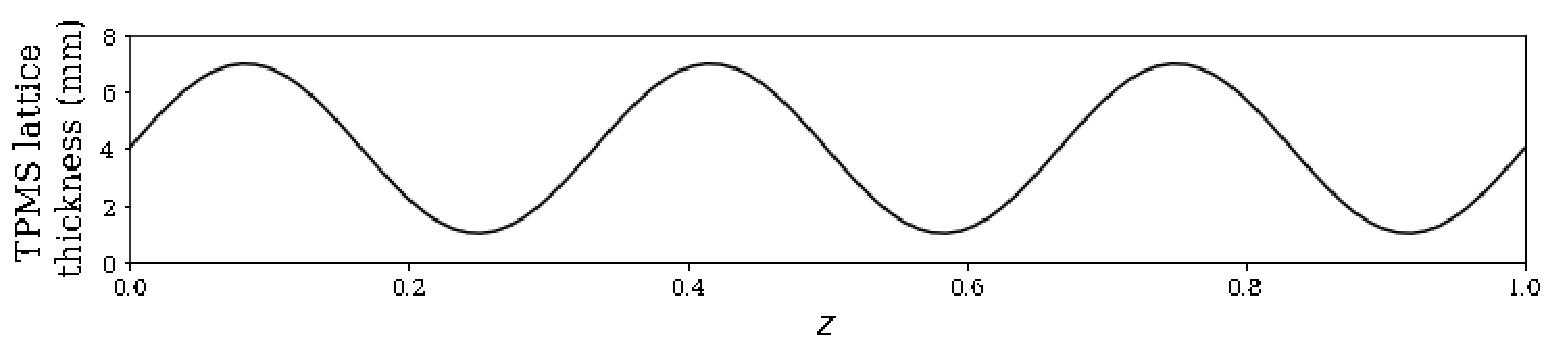}
      \subcaption{The parameter defining function
      $P$ that corresponds to the thickness of a
      TPMS-based solid body vs the $z$ coordinate}
      \label{fig:sin}
    \end{minipage}
    \caption{A heterogeneous lattice structure with
    topology based on the Schwarz P surface
    with non-linearly varying thickness}
    \label{fig:schwartz-fun}
\end{figure*}

Figure~\ref{fig:schwartz-fun} provides one more visual
comparison between
the parameter defining function $P$ and the
resulting solid model.
In this example, the thickness of the TPMS-based lattice
is controlled by the sine function:
\begin{equation}
    P(\mathbb{X}): t(z) = 3\sin{(6\pi x)} + 4,
\end{equation}
so that the thickness varies in the range $t\in[1.0,7.0]$.
The size of the lattice is $N_x=10$, $N_y=10$, $N_z=20$.

\begin{figure*}
    \centering
    \includegraphics[width=0.8\linewidth]{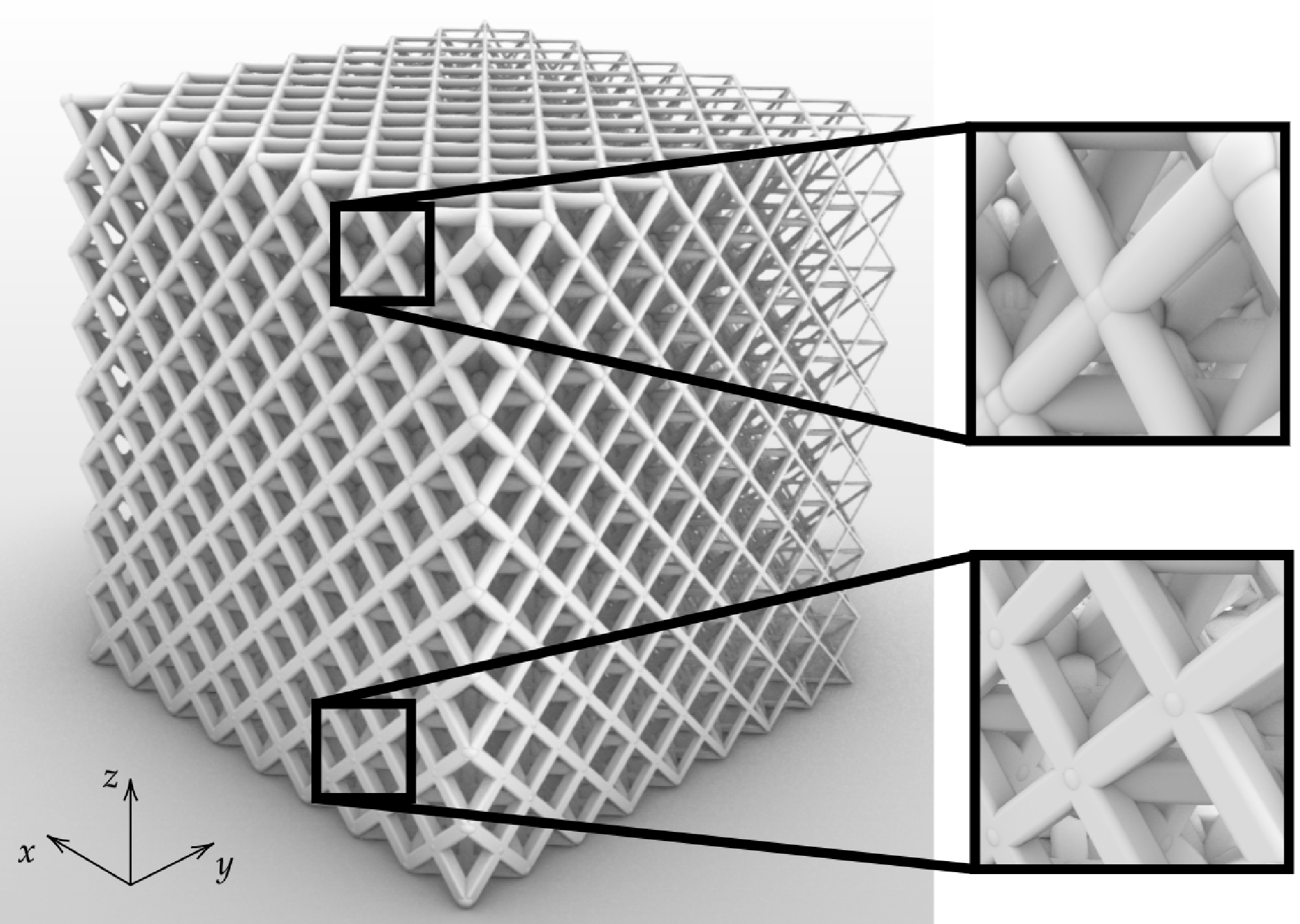}
    \caption{A heterogeneous lattice structure with the FCC
  topology with varying thickness and beam cross-section}
  \label{fig:rect-circ}
\end{figure*}

\begin{figure}
    \centering
    \begin{subfigure}[b]{\subfigw}
        \centering
        \includegraphics[width=\linewidth]{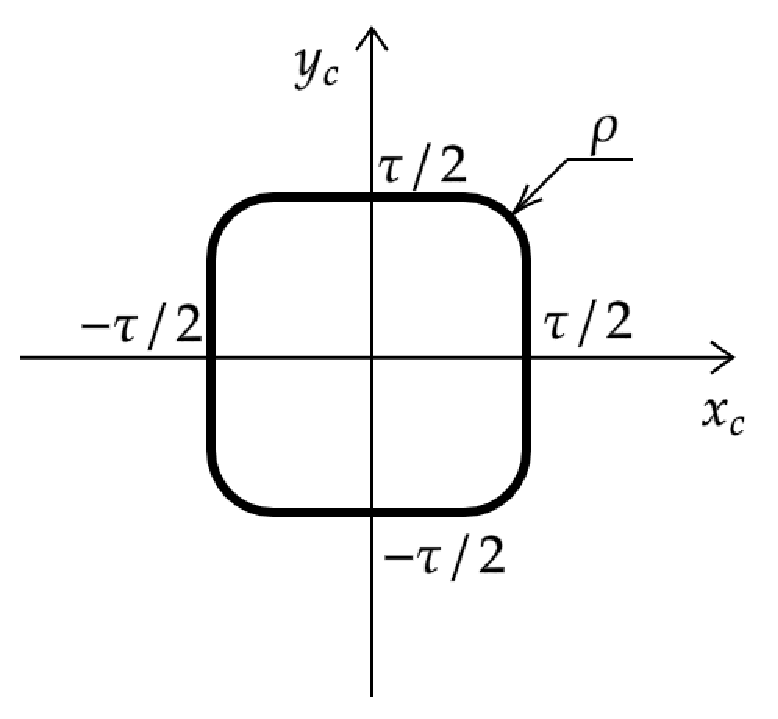}
        \caption{Cross-section of a beam in a shape of a square with rounded corners}
        \label{fig:rect-c-section}
    \end{subfigure}
    \hfill
    \begin{subfigure}[b]{\subfigw}
        \centering
        \includegraphics[width=\linewidth]{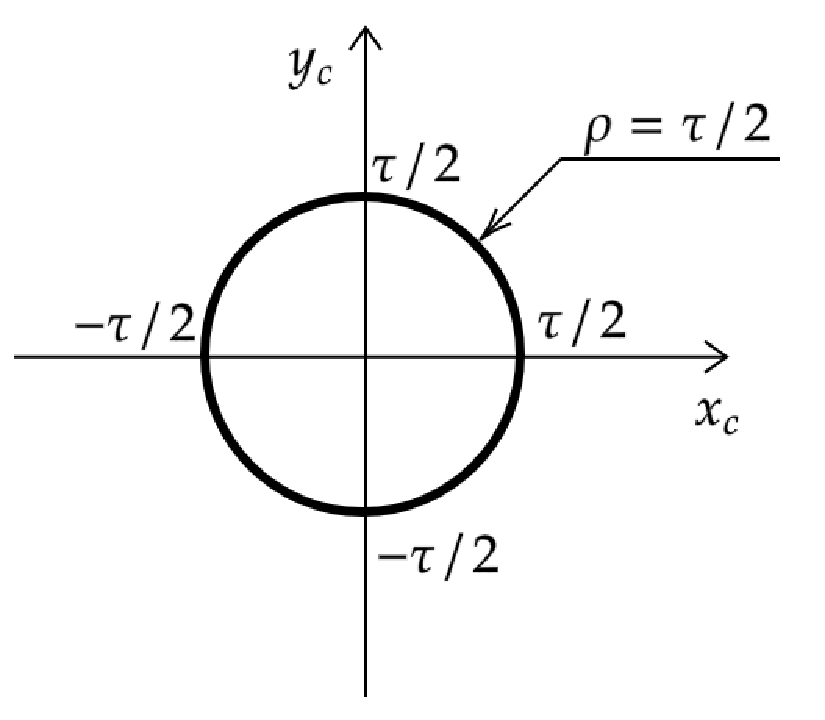}
        \caption{Cross-section of a beam in a shape of a circle}
        \label{fig:circ-c-section}
    \end{subfigure}
    \caption{Two extreme cases of the shape of the beam cross-section}
    \label{fig:c-section}
\end{figure}

A single lattice structure can have several geometric parameters varying in different
directions.
Consider the heterogeneous lattice structure illustrated in
Fig.~\ref{fig:rect-circ}\hl{,
which is modelled} with the developed software prototype.
This lattice structure has an FCC topology and has its beam size $\tau$ decreasing along the
$y$-axis linearly from 2 mm to 0.5 mm.
However, the shape of the cross-section of the beam changes along the $z$-axis,
which is uncommon in existing lattice \hl{modelling} tools.
The general shape of the beam cross-section, in this case, is a square with the side
length $\tau$ and with vertices rounded with a fillet of the radius $\rho$ as sketched
in Fig.~\ref{fig:rect-c-section}.
For this lattice, $\rho$ is set to increase \hl{linearly} from $0.2\tau$
to $0.5\tau$.
Note that $\rho=0.5\tau$ is the extreme case in which the shape of the cross-section
converges to a circle\hl{,} as sketched in Fig.~\ref{fig:circ-c-section}.
\hl{Such a transition between the two different beam cross-sections can be applied in,
for example, AM of bone implants.
It has been shown that the area moment of inertia of a beam cross-section can be used as an
indicator of mandible stiffness of the implant
patient}~\parencite{hansson2004area}\hl{.
Control over the beam cross-section and, by extent, of the area moment of inertia can
enhance the AM of bone implants that feel more natural to the patient and
have a lesser chance of being rejected by the body.}

\begin{figure}
    \centering
    \begin{subfigure}[b]{\subfigw}
        \centering
        \includegraphics[width=\linewidth]{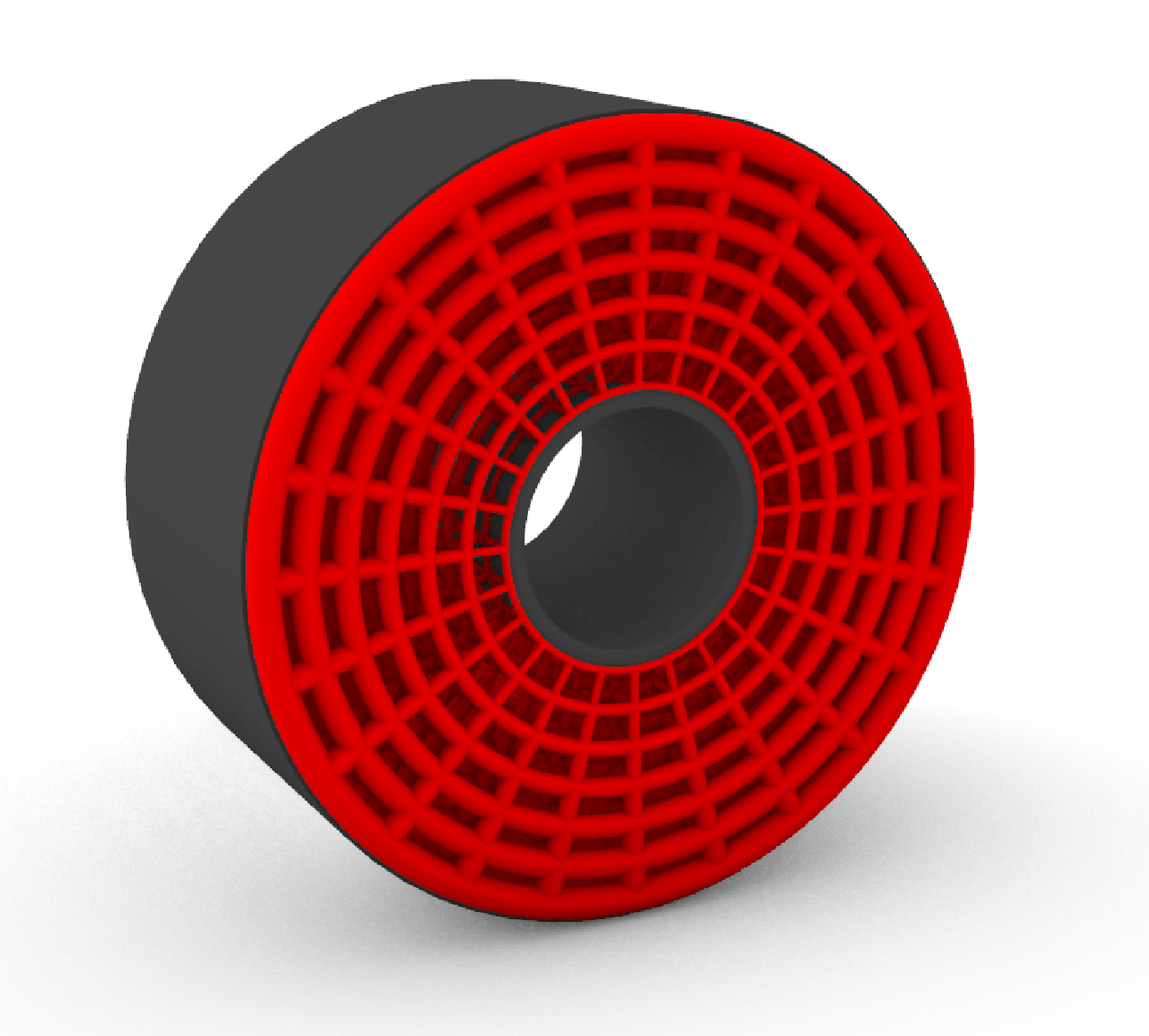}
        \caption{}
        \label{fig:tire-iso}
    \end{subfigure}
    \hfill
    \begin{subfigure}[b]{\subfigw}
        \centering
        \includegraphics[width=\linewidth]{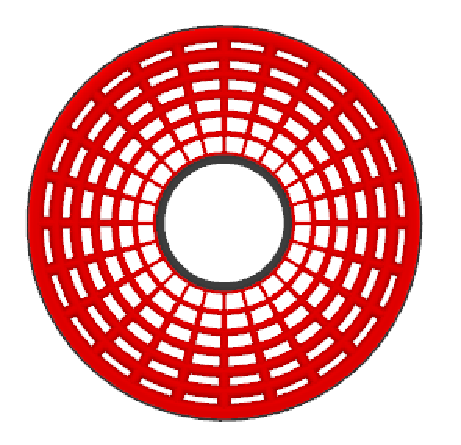}
        \caption{}
        \label{fig:tire-prof}
    \end{subfigure}
    \caption{(a) An isometric and (b) a profile view on a conformal
    heterogeneous lattice tire design modelled with the beam thickness increase
    further away from the wheel axle}
    \label{fig:tire}
\end{figure}

Similarly to Intralattice\hl{,} which was developed in previous
work~\parencite{Kurtz2015}, the proposed approach allows the \hl{modelling} of
conformal lattice structures.
Intralattice allows the \hl{modelling} of conformal lattice structures by transforming the coordinate
system to a different one\hl{,} such as the example illustrated in
Fig.~\ref{fig:intralattice-tire}.
However, in this case, the lattice density decreases further away from
the \hl{wheel's axle}.
In cylindrical coordinates, this lattice structure is homogeneous as radius $\rho$, angle $\phi$,
and $z$-axis remain constant for each unit cell.
In Cartesian coordinates, $x=\rho\cos(\phi)$ and $y=\rho\sin(\phi)$ cannot remain constant
in this case.
Moreover, the torsional shear stress increases linearly with
$\rho$~\parencite{den1961strength}.
The proposed approach allows increasing the beam thickness in $\rho$-direction so that the
lattice becomes heterogeneous even in cylindrical coordinates\hl{,} as seen in
Fig.~\ref{fig:tire}\hl{, thus illustrating a possible example of a
real-world application}.

\subsection{Computational performance}
\label{sec:comp-performance}

The testing of the developed software prototype was performed on a machine
equipped with the AMD Ryzen™ 7 3700X CPU with
3.20 GHz of clock rate, the NVIDIA® GeForce® RTX 2070 Super GPU with 8 GB of memory, 16 GB of random-access memory (RAM), a solid-state
drive (SSD) and the Linux operating system.
The mesh precision was set to be 0.1 mm.
\hl{Note that CadQuery and, by extent, the developed software prototype allow
changing the mesh precision in the settings.}
The performance of the software prototype of the proposed approach is listed in
Table~\ref{tbl:comp}.
The generation time for the beam-based lattice structures was discovered to be
1.843 s per unit cell on average (as a result of all topologies tested once
with 0.1 mm precision).
For TPMS-based lattice structure, the results are 6.453 s per unit cell on average.
\hl{Note that the difference in computational efficiency
of the geometric modelling of
both homogeneous and heterogeneous lattice structures is negligible.
This is because this work focuses mainly on the modelling of
heterogeneous lattice structures, and homogeneous lattice structures lie outside
the scope of this work.}

\begin{table*}[h!]
    \centering
    \begin{tabular}{ | m{1.55cm} | m{1.3cm} | m{1.25cm} | m{1.5cm} | m{1.5cm} | m{1.5cm} | m{1.9cm} | }
      \hline
      Topology & Number of unit cells & Figure & CPU usage range, \% & GPU usage range, MB & RAM usage, \% & Generation time, min \\ \hline
      Schwarz P & 
      1000
      &
      \ref{fig:hetero-schwartz}
      &
      65.5--104.0
      &
      388--406
      &
      2.7
      &
      122.70
      \\ \hline
      BCC & 
      720
      &
      \ref{fig:bcc-fun}
      &
      36.8--101.0
      &
      344--452
      &
      1.7
      &
      23.20
      \\ \hline
      Schwarz P & 
      2000
      &
      \ref{fig:schwartz-fun}
      &
      54.7--105.3
      &
      390--533
      &
      2.2
      &
      121.30
      \\ \hline
    \end{tabular}
    \caption{The performance metrics of \hl{the modelling}
    with the proposed approach
    }
    \label{tbl:comp}
\end{table*}

The resulting model can be saved as an STL file
and as a STEP file.
\hl{The STL and STEP export are} made possible by the support of STL and
STEP by Open CASCADE.
\hl{Notably, FLatt Pack and MSLattice do not support export to STEP.
The STEP file format is rarely used in AM itself but can be used as an input for
a CAE simulation in software such as Ansys}~\parencite{ansys}.

The output STL files were successfully imported into slicing
tools such as Ultimaker Cura~\parencite{cura} and
Preform~\parencite{preform}.
As a potential future feature, a pivot from ASCII STL to binary STL could be made
to decrease the size of the output files.
CadQuery supports only ASCII STL files, but Open CASCADE
supports binary STL files.
Thus, building an extra method for CadQuery to be able to export
binary STL files will be a focus for further \hl{optimisation}.
Also, as of now, the proposed approach is packed into an
importable library, but no significant effort was made to provide a proper
GUI for the proposed \hl{method}.
While this was deemed acceptable for the scope of this work to provide
a proof of concept, it can be more appealing for the actual
lattice designers to have a proper GUI.

While the performance of the developed software prototype is suitable for an MVP,
for future improvements\hl{, it is
essential} to consider an enhanced geometric \hl{modelling} approach with
access to the low-level GMK functionality~\parencite{Xu2009}.
It might involve developing a novel GMK to enable functionality not provided by Open
CASCADE.
\hl{The development of a GMK, however, is outside the scope of}
the proposed MVP, as a GMK is often being developed
by a substantial team of mathematicians and programmers over the course of several
years~\parencite{golovanov2014geometric}.
At the same time, having a GMK highly tuned to the scope of the proposed work
does not solve the immediate problem of \hl{the modelling}
of heterogeneous lattice structures.

The performance of the software prototype is decent
for \hl{the modelling of a} geometrically complex heterogeneous
lattice structure.
The software prototype proved itself slower than geometric \hl{modelling}
approaches that deal
with linearly varying geometric parameters and \hl{homogeneous lattice
structures}.
\hl{The computation efficiency is mainly set back}
because, in the proposed method\hl{,} every unit cell needs its
geometric parameters calculated prior \hl{to
the rendering,} and every instance of the unit cell is unique.
Such \hl{behaviour} is expected to be slower compared to
\hl{the modelling} of homogeneous lattice structures.
Further investigation is required to increase the performance of the software prototype
for any future work involving it.
Moreover, the performance gap between the generation of beam-based lattices and
TPMS lattices is noticeable, which is explained \hl{mainly}
by the higher geometrical complexity of the latest.

It is also noticeable that the software prototype does not make much use of the
GPU and almost solely relies on the CPU.
Appropriate \hl{utilisation} of the GPU is crucial for \hl{a geometric modelling}
tool operating with lattice structures, as GPUs are \hl{generally} better
suited for dealing
with large amounts of parallel tasks.
\hl{Specific changes must} be made to the GPU \hl{utilisation}
process to make the \hl{modelling} process more efficient.

There is evidence that considering the bio-inspired nature of lattice structures
in general, the geometric \hl{modelling} approach suitable for their
\hl{generation could utilise} bio-inspired algorithms as well~\parencite{Letov2020}.

\section{Conclusions and future research}
\label{sec:conclusions}
This work presented \hl{a geometric modelling approach that is based on F-rep
and} expands the freedom of designing heterogeneous lattice structures.
Functionally controllable
topology parameters such as thickness (for TPMS) and beam diameter (for beam-based lattices)
can be \hl{modelled} and controlled by the proposed approach.
\hl{The proposed method} was implemented in a software prototype and tested.

For future research, improvements to the software prototype are proposed.
These improvements include computational \hl{optimisation}
based on the results obtained in Section~\ref{sec:comp-performance}, providing a proper
GUI for the usage of the tool and other improvements.
\hl{A proper GUI would allow the proposed approach to be integrated into a generic
design workflow with the most immediate goal of collecting user feedback.
This feedback can be then used to improve the GUI and enhance the software performance.
}

\hl{The methodology} is also proposed to be improved in future research.
This work mainly focuses on the control of the geometric parameters $P$, but further
investigation on controlling the topology $T$ \hl{is} required.
\hl{This control} should enable a smooth transition between topologies, including TPMS.
\hl{The design spaces presented in this work are limited to being
standard and linear, even though the proposed approach can model conformal
lattice structures by transforming the coordinates into another
3D coordinate system.
Allowing control over }$T$\hl{ should assist with the modelling
of lattice structures with nonstandard and non-linear design spaces.}
The modelling of multi-scale lattices is proposed to be achieved by introducing $P_l$ and $T_l$
for every level of detail $l$.

Potential applications of the proposed approach for enhancing topology
\hl{optimisation} techniques are proposed to be investigated for future research.

\part{Acknowledgements}
This research work is supported by National Sciences and Engineering Research
Council of Canada Discovery Grant RGPIN-2018-05971 \hl{and by
the McGill Engineering Doctoral Award}.

\part{Conflict of interest statement}
The authors declared no potential conflicts of interest with respect
to the research, authorship, and/or publication of this article.

\printbibliography

\clearpage

\appendix

\section{Functional representation of skeletal graphs of lattice topologies}
\label{sec:tables}
This section includes function representation of skeletal graphs of lattice
topologies \hl{following} the F-rep method proposed in
Section~\ref{sec:frep}.
The skeletal graphs define only the frame of the lattice of zero-thickness.
Setting the thickness $t>0$ allows \hl{the modelling} of solid bodies as described in
Section~\ref{sec:implementation}.

Table~\ref{tbl:topologies} lists beam-based topologies
inspired by the cubic crystal system
along with their topology-defining functions $T$.
The topology-defining function $T$ in this case is defined for
$x,y,z\in[0,u]$.
Note that in Table~\ref{tbl:topologies} the unit cell is assumed to be
cubic with the side $u$.
The notations of the form $x\in\{a,b\}$ that are used
in this work denote the union of two parallel lines
with $x=a$ and $x=b$.
In other words,
\begin{equation}
    x\in\{a,b\} \Leftrightarrow \lnot [a,b] \cup (a,b).
\end{equation}

\begin{table*}[h!]
    \centering
    \begin{tabular}{ | c | m{6cm} | m{3cm} | }
      \hline
      Topology & Topology defining function, $T$ & Figure \\ \hline
      Simple cubic
      &
      $
      \left[ 
        \begin{gathered} 
          x \in \{0, u\}, y \in \{0,u\}, \\ 
          y \in \{0, u\}, z \in \{0,u\}, \\
          x\in= \{0, u\}, z \in \{0,u\}.
        \end{gathered} 
      \right.
      $
      &
      \ref{fig:cubic}
      \\ \hline
      BCC
      &
      $
      \left[ 
      \begin{gathered} 
        x=y=z, \\ 
        -x+u=y=z, \\
        x=y=-z+u, \\ 
        -x+u=y=-z+u. 
      \end{gathered}
      \right.
      $
      &
      \ref{fig:bcc}
      \\ \hline
      S-FCC
      &
      $
      \left[ 
      \begin{gathered} 
        x\in\{0,u\}, z \in \{y, -y + u\}, \\
        y \in \{0, u\}, z \in \{x, - x + u\}.
      \end{gathered}
      \right.
      $
      &
      \ref{fig:sfcc}
      \\ \hline
      FCC
      &
      $
        \text{S-FCC }\cup\
        \left\{
            \begin{gathered}
                z \in \{0, u\}, \\
                y \in \{x, - x + u\}.
            \end{gathered}
        \right.
      $
      &
      \ref{fig:fcc}
      \\ \hline
      BCCz
      &
      $
        \text{BCC }\cup\
        \left\{
            \begin{gathered}
                x \in \{0, u\}, \\
                y \in \{0, u\}.
            \end{gathered}
        \right.
      $
      &
      \ref{fig:bccz}
      \\ \hline
      FCCz
      &
      $
        \text{FCC }\cup\
        \left\{
            \begin{gathered}
                x \in \{0, u\}, \\
                y \in \{0, u\}.
            \end{gathered}
        \right.
      $
      &
      \ref{fig:fccz}
      \\ \hline
      S-FCCz
      &
      $
        \text{S-FCCz }\cup\
        \left\{
            \begin{gathered}
                x \in \{0, u\}, \\
                y \in \{0, u\}.
            \end{gathered}
        \right.
      $
      &
      \ref{fig:sfccz}
      \\ \hline
      FBCC
      &
      $
        \text{BCC }\cup\text{ FCC}.
      $
      &
      \ref{fig:fbcc}
      \\ \hline
      S-FBCC
      &
      $
        \text{BCC }\cup\text{ S-FCC}.
      $
      &
      \ref{fig:sfbcc}
      \\ \hline
      S-FBCCz
      &
      $
        \text{BCC }\cup\text{ S-FCCz}.
      $
      &
      \ref{fig:sfbccz}
      \\ \hline
    \end{tabular}
    \caption{Various beam-based topologies
    inspired by the cubic crystal system
    supported by the developed approach}
    \label{tbl:topologies}
\end{table*}

Tables~\ref{tbl:diamond}, \ref{tbl:rco}, and \ref{tbl:tcube} describe the diamond,
rhombicuboctahedron, and truncated cube topologies, respectively.
The rhombicuboctahedron and truncated
cube topologies require an additional truncation
parameter $t\in[0, 0.5u]$ which sets the size of
truncation.

\begin{table*}[h!]
    \centering
    \begin{tabular}{ | c | m{9cm} | m{1.5cm} | }
      \hline
      Topology & Topology defining function, $T$ & Figure \\ \hline
      Diamond
      &
      for $z\in[0,0.25u]$:

      $
      \left[
        \begin{gathered}
            -x+u=y=z, \\
            x-0.5u=-y+0.5u=z, \\
            -x+0.5u=y-0.5u=z, \\
            x=-y+u=z;
        \end{gathered}
      \right.
      $

      for $z\in[0.25,0.5u]$:

      $
      \left[
        \begin{gathered}
            -x+0.75u=-y+0.25u=z-0.25u, \\
            x-0.75u=y-0.25u=z-0.25u, \\
            x-0.25u=y-0.75u=z-0.25u, \\
            -x+0.25u=-y+0.75u=z-0.25u;
        \end{gathered}
      \right.
      $

      for $z\in[0.5,0.75u]$:

      $
      \left[
        \begin{gathered}
            -x+0.5u=y=z-0.5u, \\
            x=-y+0.5u=z-0.5u, \\
            -x+u = y-0.25u = z-0.5u, \\
            x-0.5u=-y+u=z-0.5u;
        \end{gathered}
      \right.
      $

      for $z\in[0.75,u]$:

      $
      \left[
        \begin{gathered}
            -x+0.25u=-y+0.25u=z-0.75y, \\
            x-0.25u=y-0.25u=z-0.75u, \\
            -x+0.75u=-y+0.75u=z-0.75u, \\
            x-0.75u=y-0.75u=z-0.75u.
        \end{gathered}
      \right.
      $
      &
      \ref{fig:diamond}
      \\ \hline
    \end{tabular}
    \caption{The diamond topology
    supported by the developed approach}
    \label{tbl:diamond}
\end{table*}

\begin{table*}[h!]
    \centering
    \begin{tabular}{ | m{2.5cm} | m{8cm} | m{1.5cm} | }
      \hline
      Topology & Topology defining function, $T$ & Figure \\ \hline
      Rhombicubo-\newline ctahedron
      &
      $
      \left[
        \begin{gathered}
            x\in\{t,u-t\}, \left[\begin{gathered}
                -y+t=z,\\
                -y-u+t=z,\\
                y=z-u+t,\\
                -y+u=z-u+t;
                \end{gathered}
            \right.
            \\
            y\in\{t,u-t\}, \left[\begin{gathered}
                x=z-u+t,\\
                -x+u=z-u+t,\\
                -x+t=z,\\
                x-u+t=z;
                \end{gathered}
            \right.
            \\
            z\in\{t,u-t\}, \left[\begin{gathered}
                -x+t=y,\\
                x-u+t=y,\\
                x=y-u+t,\\
                x-u+t=-y+u;
                \end{gathered}
            \right.
            \\
            x\in\{t,u-t\}, y\in\{0,u\}, z\in[t,u-t];
            \\
            x\in\{0,u\}, y\in\{t,u-y\}, z\in[t,u-t];
            \\
            x\in\{t,u-t\}, z\in\{0,u\}, y\in[t,u-t];
            \\
            x\in\{0,u\}, z\in\{t,u-y\}, y\in[t,u-t];
            \\
            y\in\{t,u-t\}, z\in\{0,u\}, x\in[t,u-t];
            \\
            y\in\{0,u\}, z\in\{t,u-y\}, x\in[t,u-t];
            \\
            z\in\{t,u-t\}, \left[\begin{gathered}
                -x+t=y, \\
                x-u+t=y, \\
                x=y-u+t, \\
                x-u+t=-y+u.
                \end{gathered}
            \right.
        \end{gathered}
      \right.
      $
      &
      \ref{fig:rco}
      \\ \hline
    \end{tabular}
    \caption{Rhombicuboctahedron topology
    supported by the developed approach}
    \label{tbl:rco}
\end{table*}

\begin{table*}[h!]
    \centering
    \begin{tabular}{ | m{2.5cm} | m{8cm} | m{1.5cm} | }
      \hline
      Topology & Topology defining function, $T$ & Figure \\ \hline
      Truncated cube
      &
      $
      \left[
        \begin{gathered}
            x\in\{0,u\}, \left[\begin{gathered}
                -y+t=z, \\
                y-u+t=z, \\
                y=z-u+t, \\
                -y+u=z-u+t;
                \end{gathered}
            \right.
            \\
            y\in\{0,u\}, \left[\begin{gathered}
                -x+t=z, \\
                x-u+t=z, \\
                x=z-u+t, \\
                x-u+t=-z+u;
                \end{gathered}
            \right.
            \\
            z\in\{0,u\}, \left[\begin{gathered}
                -x+t=y, \\
                x-u+t=y, \\
                x=y-u+t, \\
                -x+u=y-u+t;
                \end{gathered}
            \right.
            \\
            x\in[t,u-t], \left[\begin{gathered}
                x\in\{0,u\}, \\
                z\in\{0,u\};
                \end{gathered}
            \right.
            \\
            y\in[t,u-t], \left[\begin{gathered}
                x\in\{0,u\}, \\
                z\in\{0,u\};
                \end{gathered}
            \right.
            \\
            z\in[t,u-t], \left[\begin{gathered}
                x\in\{0,u\}, \\
                y\in\{0,u\}.
                \end{gathered}
            \right.
        \end{gathered}
      \right.
      $
      &
      \ref{fig:t-cubic}
      \\ \hline
    \end{tabular}
    \caption{Truncated cube topology
    supported by the developed approach}
    \label{tbl:tcube}
\end{table*}

Table~\ref{tbl:tpms-topologies} covers 
functions $T$ that define supported TPMS topologies.
Note that $T$ these surfaces are
approximations of their exact form~\cite{gandy1999exact}.

\begin{table*}[h!]
    \centering
    \begin{tabular}{ | m{2cm} | p{8cm} | m{2cm} | }
      \hline
      Topology & Topology defining function, $T$ & Figure \\ \hline
      Gyroid
      &
      $
      \sin{x}\cos{y}+\sin{y}\cos{z}+\sin{z}\cos{x}=0
      $
      &
      \ref{fig:gyroid}
      \\ \hline
      Schwarz P surface
      &
      $
      \cos{x}+\cos{y}+\cos{z}=0
      $
      &
      \ref{fig:schwartz}
      \\ \hline
      Schwarz D surface
      &
      $\cos{x}\cos{y}\cos{z}-\sin{x}\sin{y}\sin{z}=0$
      &
      \ref{fig:schwartz-d}
      \\ \hline
    \end{tabular}
    \caption{Various TPMS-based topologies
    supported by the developed approach}
    \label{tbl:tpms-topologies}
\end{table*}

\end{document}